\documentclass[apj, numberedappendix]{emulateapj}
\usepackage{epsfig}
\usepackage{graphicx}

\usepackage{amsfonts}
\usepackage{amsmath}
\usepackage{mathrsfs}
\usepackage{amssymb}
\usepackage{rotating}
\usepackage{float}

\usepackage{braket}

\usepackage{natbib}
\usepackage{hyperref}

\usepackage{color}
\usepackage{ulem}
\usepackage{xspace}
\usepackage{cancel}
\definecolor{lgray}{gray}{0.85}

\newcommand{\lya} {Ly$\alpha$\xspace}
\newcommand{\lyb} {Ly$\beta$\xspace}
\newcommand{\lcont}{$\mathcal{L_{\rm cont}}$~}
\newcommand{\lpdf}{$\mathcal{L_{\rm PDF}}$~}

\newcommand{\heii}{\ion{He}{2}\xspace}

\def\bea{\begin{eqnarray}}
\def\eea{\end{eqnarray}}

\newcommand*\diff{\mathop{}\!\mathrm{d}}

\begin{document}

\title{Joint Bayesian Estimation of Quasar Continua and the Lyman-Alpha Forest Flux Probability Distribution Function}

\author{Anna-Christina Eilers\altaffilmark{1,2}\altaffilmark{*}, Joseph F. Hennawi\altaffilmark{1,3}, Khee-Gan Lee\altaffilmark{1,4}}
\altaffiltext{*}{email: eilers@mpia.de}
\altaffiltext{1}{Max-Planck-Institute for Astronomy, K\"onigstuhl 17, 69117 Heidelberg, Germany; eilers@mpia.de}
\altaffiltext{2}{International Max Planck Research School for Astronomy \& Cosmic Physics at the University of Heidelberg}
\altaffiltext{3}{Physics Department, University of California, Santa Barbara, CA 93106-9530, USA}
\altaffiltext{4}{Lawrence Berkeley National Laboratory, Berkeley, CA 94720, USA}

\slugcomment{Draft Version of \today}
\shortauthors{Eilers et al.}

\begin{abstract}

We present a new Bayesian algorithm making use of Markov Chain Monte
Carlo sampling that allows us to simultaneously estimate the
unknown continuum level of each quasar in an ensemble of
high-resolution spectra, as well as their common probability
distribution function (PDF) for the transmitted \lya forest flux. This
fully automated PDF regulated continuum fitting method models the
unknown quasar continuum with a linear Principal Component Analysis
(PCA) basis, with the PCA coefficients treated as nuisance
parameters. The method allows one to estimate parameters governing the
thermal state of the intergalactic medium (IGM), such as the slope of
the temperature-density relation $\gamma-1$, while marginalizing out
continuum uncertainties in a fully Bayesian way. Using realistic mock
quasar spectra created from a simplified semi-numerical model of the
IGM, we show that this method recovers the underlying quasar continua
to a precision of $\simeq7\%$ and $\simeq10\%$ at $z=3$ and $z=5$,
respectively. Given the number of principal component spectra,
this is comparable to the underlying accuracy of the PCA model itself.
Most importantly, we show that we can achieve a nearly unbiased
estimate of the slope $\gamma-1$ of the IGM temperature-density relation
with a precision of $\pm8.6\%$ at $z=3$, $\pm6.1\%$ at
$z=5$, for an ensemble of ten mock high-resolution quasar
spectra. Applying this method to real quasar spectra and
comparing to a more 
realistic IGM model from hydrodynamical simulations would enable precise measurements of the thermal
and cosmological parameters governing the IGM, albeit with somewhat
larger uncertainties given the increased flexibility of the model. 

\end{abstract}
    

\maketitle

\section{Introduction}

The imprint of Lyman-$\alpha$ (Ly$\alpha$) forest absorption lines
observed on the spectra of distant quasars, caused by residual neutral
hydrogen along the line of sight in a mostly ionized intergalactic
medium (IGM), has become an important tool for constraining cosmology
and the IGM at high redshift $2\lesssim z\lesssim 6$ \citep[see e.g.][]{Croft1998, Weinberg2003, Zaldarriaga2003, Meiksin2009_review, BeckerReview2015, MortlockReview2015, McQuinnReview2015}. The observed
absorption pattern traces density fluctuations of the IGM along the
filamentary structure of the cosmic web that arise due to
gravitational instability in an universe dominated by cold dark matter \citep[see e.g.][]{Cen1994, MiraldaEscude1996, Dave1999}.

An important characteristic of the fluctuating
IGM is a tight relationship between the temperature $T$ and the density
contrast $\Delta$ of the cosmic gas \citep{Hui1997, McQuinn2015}.
This `equation of state' 
is controlled by the
interplay of two mechanisms: photoionization heating
by the ultraviolet background (UVB)
radiation
and adiabatic cooling due to the expansion of the universe. 
Assuming a power-law relationship for the temperature-density relation we can express
the equation of state as 
\begin{align}
T(\Delta) = T_0 \Delta^{\gamma -1}, \label{eq:T-Drelation}
\end{align}
where $\Delta = \rho/\bar{\rho}$, $\rho$ is the matter density field,
and $\bar{\rho}$ denotes the mean density of the universe.  The
temperature at mean density is denoted by $T_0$ and the parameter
$\gamma-1$ describes the slope of this temperature-density relation 
indicating whether overdense regions in the universe are hotter than
underdense voids (i.e. $\gamma>1.0$) or vice versa (i.e. inverted
temperature-density relation, $\gamma<1.0$). 

Statistical properties of the transmitted flux in
the Ly$\alpha$ forest, such as the probability distribution function
(PDF) or the line-of-sight power spectrum, provide information about the underlying physics
governing the IGM and hence the thermal evolution of the universe 
\citep[see e.g.][]{McDonald2000, Weinberg2003, Meiksin2009_review}. The \lya flux PDF --- 
although dependent on all thermal
parameters --- is particularly sensitive to the value of the slope parameter $\gamma$ of the temperature-density relation,
that provides valuable insights into the thermal state of the IGM, and
thus the PDF represents a useful tool to obtain constraints on it \citep{Jenkins1991}.
Several studies using this statistical approach to infer $\gamma$ have found
evidence for an inverted,
i.e. $\gamma<1.0$ \citep[see e.g.][]{Becker2007, Viel2009, Bolton2008, Rorai2016}
or isothermal, i.e. $\gamma\approx1.0$ \citep[see e.g.][]{Calura2012, Garzilli2012}, 
temperature-density relation of the IGM. However, an inverted
temperature-density relation contrasts with the theoretically
predicted value of $\gamma\approx 1.6$ for a post-reionization IGM \citep[see e.g.][]{Hui1997, HuiHaiman2003, McQuinn2015}. Furthermore, it has proven to be difficult to explain an inverted temperature-density relation. Blazar heating has been suggested as a possible mechanism to cause an inverted relation \citep[e.g.][]{Puchwein2012, Broderick2012, Chang2012, Pfrommer2012}, as well as additional heating of the IGM due to \heii reionization \citep[e.g.][]{Bolton2008, Furlanetto2008, McQuinn2009, MeiksinTittley2012, Compostella2013}. However, both mechanisms fail to reproduce an inverted temperature-density relation at typical densities around $z\approx 3$.

Other attempts to constrain $\gamma$ from the
the \lya flux PDF, did not find
evidence for an inverted temperature-density relation \citep[see e.g.][]{Rollinde2013, LeeHennawi2015}. 
The one-dimensional power spectrum has also been applied to obtain constraints on $\gamma$ \citep[e.g.][]{McDonald2000, Zaldarriaga2001, Zaldarriaga2003}. These studies favor a slope parameter of $\gamma > 1.0$, which is more consistent with the canonical value of $\gamma\approx1.6$. Other techniques to infer $\gamma$ such as wavelet decomposition \citep[e.g.][]{Theuns2000, Lidz2010} or the measurement of Doppler parameters and column densities of individual \lya forest absorbers \citep[e.g.][]{Rudie2012, Bolton2014, McDonald2001, Schaye2000} lead to similar estimates of $\gamma\approx1.3-1.6$.

The reason for the differences between the estimates of the slope parameter
$\gamma$ of the temperature-density relation are still debated in the
literature, since the various techniques have different sources of
systematic uncertainties. In this paper we will focus on the \lya flux
PDF, which seems to be giving the widest spread of values for the
estimate of $\gamma$. One challenge and the biggest source of
uncertainty when working with the PDF is that it requires precise
estimates of the quasar continuum level for each quasar spectrum,
since imprecise continuum fitting can result in biases \citep[see e.g.][]{Desjacques2007, Lee2012}.  \citet{Kim2007} show
that different continuum fitting strategies applied to the same data
set can result in significant differences in the resulting PDF and
hence varying estimates for $\gamma$. \citet{FaucherGiguere2008} show
that the magnitude of the bias in the continuum estimation increases
with redshift, since it becomes more and more difficult to estimate
the unabsorbed continuum level for quasars at $z\gtrsim 4$ due to
increased \lya absorption. 
\citet{Lee2012} also pointed out that the quasar continuum fit is dependent on the underlying model of the IGM. Thus depending on the true underlying value of 
$\gamma$ one is more likely to over- or underestimate the quasar continua. A larger (smaller) value of $\gamma$ leads to
an under (over) -estimation of the continuum level. 
In order to avoid likely biases due to continuum fitting by hand, \citet{Lee2012_MeanFlux} introduced mean-flux-regulated principal component analysis (PCA) continuum fitting, where
PCA fitting is carried out on wavelengths longer than the \lya emission
line in order to provide a prediction for the shape of the \lya
forest continuum. The slope and amplitude of this continuum prediction is then corrected using external constraints
for the \lya forest mean flux. 

A similar approach for handling continuum uncertainties in the estimation of the flux PDF has been presented by \citet{Rorai2016} recently, who reduced the sensitivity of the PDF to these uncertainties by calculating the flux PDF from a `regulated' flux level, defined as the transmitted flux divided by the $95$th percentile of the flux distribution within a spectral region of $10$~Mpc~$h^{-1}$. The advantage of regulating the transmitted flux by the $95$th percentile of the flux distribution compared to the mean flux is, that the $95$th percentile of the flux falls near the peak of the flux PDF for all IGM models and is therefore less noisy than the mean flux, which falls in a flux interval with low probability. 

In this paper we present a new Bayesian algorithm making use of a Markov Chain
Monte Carlo (MCMC) sampling that allows us to simultaneously estimate the
unknown continuum of each quasar in an ensemble of high-resolution spectra as well as
their common \lya forest flux PDF. This fully automated PDF regulated continuum fitting
method models the unknown quasar continuum with a PCA with the
coefficients of the principal components treated as nuisance parameters. This method
allows us to estimate parameters governing the thermal state of the IGM, such as the
slope parameter of the temperature-density relation $\gamma$, while
marginalizing out continuum uncertainties in a fully
Bayesian way. We are thus also able to investigate any degeneracies between the
uncertainties in the continuum estimation and the thermal properties
of the IGM.

The paper is structured as follows. In \S~\ref{sec:mock_spectra} we
describe our method for generating mock data of high-resolution quasar
spectra with a lognormal model for \lya forest absorption required to
develop our analysis algorithm. In \S~\ref{sec:pdf_regulation} we introduce
the likelihood function of the
PDF of the transmitted \lya forest flux and explain how it can be
used to simultaneously estimate the quasar continuum level and the
thermal properties of the IGM governing the shape of the PDF. 
In \S~\ref{sec:results} we
present the results of our analysis at two different redshifts, $z=3$
and $z=5$, and study the degeneracies between thermal parameters of the
IGM and continuum uncertainties. We discuss and summarize our key results in \S~\ref{sec:discussion} and 
\S~\ref{sec:summary}.

\section{Mock Quasar Spectra with Ly$\alpha$ Forest Absorption}\label{sec:mock_spectra} 

In this section we describe a method to generate matter density
fields in order to create mock \lya forest absorption spectra. This
\lya forest model can then be used to generate realistic mock quasar spectra by
multiplying the absorption field with a quasar continuum level.

\subsection{A Semi-Analytic Model for Ly$\alpha$ Forest Absorption}\label{sec:mock_lya_forest} 

The fluctuating Gunn-Peterson approximation \citep[FGPA, see e.g.][]{Weinberg1998, Croft1998, Croft1999}
provides a relationship between the observed flux in the \lya forest
and the underlying matter density distribution. It represents a
valuable tool for generating Ly$\alpha$ absorption spectra from cosmic
matter density fields, which we will use to generate mock quasar
spectra. Assuming that the intergalactic gas is in photoionization
equilibrium 
and follows the temperature-density
relation in eqn.~(\ref{eq:T-Drelation}), the optical depth $\tau$ of
the IGM scales with the matter density contrast as
\begin{align}
\tau \propto \Delta^{2-0.7(\gamma-1)}. \label{eq:fgpa}
\end{align}
The normalized transmitted flux $F$ in the \lya forest is related to the intervening optical depth of the photoionized gas by
\begin{align}
F=\exp(-\tau). \label{eq:exp_tau}
\end{align}
Combining eqn.~(\ref{eq:fgpa}) and eqn.~(\ref{eq:exp_tau}) leads to a relationship between the transmitted flux $F$ and the matter density contrast $\Delta$.

\citet{McDonald2006_mockforest} provide a semi-analytic model to
generate absorption fields for the \lya forest from matter density
fields based on the lognormal model introduced by \citet{Bi1992} and
\citet{BiDavidsen1997}. The goal of this semi-analytic model is to
generate realistic flux fields $F$ that approximately reproduce both
the observed power spectrum as a function of wave number $k$ and
redshift $z$ and the observed PDF of the \lya forest.  We introduce
small modifications to this model in order to accommodate different
values for $\gamma$ in the temperature-density relation, and summarize
its main features below.

The generation of each $F$ field
for the $i$th quasar spectrum begins by creating an initial Gaussian random field $\delta_{i,0}$ representing the underlying 
dark matter density contrast with a power spectrum given by:  
\begin{align}
P_{\delta}(k) = \frac{1+\left(\frac{0.01 \rm s/km^{-1}}{k_0}\right)^{\nu}}{1+\left(\frac{k}{k_0}\right)^{\nu}}\exp\left[-\left(k R_{\delta}\right)^2\right], \label{eq:ps}
\end{align}
with values of $k_0=0.001$ s/km, $\nu=0.7$, $R_{\delta}=5.0$ km/s, that were chosen to approximately reproduce the dependence on wave number $k$ in the observed flux power spectrum of the Ly$\alpha$ forest.
An evolution of the amplitude $A(z)$ of the density fluctuations with redshift is introduced by the transformation
\begin{align}
\delta_i = A(z_i) \delta_{i,0}
\end{align} 
with  
\begin{align}
A^2(z_i) = 58.6\left(\frac{1+z_i}{4}\right)^{-2.82}. 
\end{align}
The values for the amplitude and exponent here were chosen such that the final flux power spectrum would evolve like the observed one. 
In order to obtain the density field $n_i$ of the baryonic diffuse matter of the IGM we do not use the squared lognormal transformation that has been used in
previous work \citep[see e.g.][]{Bi1992, BiDavidsen1997, McDonald2006_mockforest}, 
but instead use a lognormal transformation with an exponent of $2-\beta$ with $\beta = 0.7(\gamma-1)$.
This step introduces the dependence on the slope parameter $\gamma$ of the temperature-density relation of the IGM that we are interested in. 
Hence the transformation of the density field $n_i$ is computed as
\begin{align}
n_i=n_{0}\left[\exp\left(\delta_i-\left\langle\frac{\sigma_i^2}{2}\right\rangle\right)\right]^{2-0.7(\gamma-1)}, 
\end{align}
where $n_0$ is the mean number density of the IGM.
The factor $\sigma_i^2$ is computed from the input density power spectrum in eqn.~(\ref{eq:ps})
and fixes the mean of the lognormal field to unity. 

After the transformation the $n_i$ field is smoothed
with a Gaussian filter with standard deviation $\sigma_{\tau}=20$ km s$^{-1}$. Note that the quasar spectra that we will consider in this work have a resolution comparable to the achievable resolution of echelle spectrographs (see \S~\ref{sec:noise} for details), which is higher than the here applied Gaussian filter. However, since this smoothing is applied to the $n_i$ field and not to the flux $F$ itself, it accounts for the effects that give rise to the velocity width of the \lya forest lines, such as thermal broadening, Jeans pressure smoothing \citep[see][]{GnedinHui1998, rorai2013, Kulkarni2015, Onorbe2016}, or peculiar velocities and velocities due to the Hubble flow.
We then multiply with a redshift evolution factor of $0.374[(1+z_i)/4]^{5.1}$ (the form of this factor is again chosen to reproduce the observations) to produce a field $\tau$. The transmitted flux can then be calculated with eqn.~(\ref{eq:exp_tau}).

Note that this simple toy model of the IGM only accommodates a
dependency on the thermal parameter $\gamma$ and ignores dependencies
on the temperature $T_0$ or the pressure smoothing scale $\lambda_P$.
However, \citet{Bolton2008} used hydrodynamical simulations to show
that the \lya flux PDF is mostly sensitive to $\gamma$, whereas the
other thermal parameters introduce only relatively modest changes to
its shape.  If we would apply our method to real quasar spectra the
simple model adopted here would clearly be insufficient, but rather we
would use hydrodynamical simulations that accommodate dependencies on
all thermal parameters. However, the main advantage of this model is
that it is fast and a large grid of models can be generated
quickly. We will comment more on this later and argue that future
augmentation of the parameter space of the PDF should be easy to
incorporate.

We re-scale the optical depth in order to obtain values for the mean
flux of our mock \lya forest spectra that are consistent with the measurements
of \citet{FaucherGiguere2008} at redshift $z=3$, i.e. $\langle F\rangle\approx 0.680$. 
For an estimate of the mean flux at $z=5$ we use the
fitting formula presented in \citet{Onorbe2016}, because the
measurements of the mean flux do not extend past $z > 4.85$, and
obtain $\langle F\rangle \approx 0.189 $ at redshift $z=5$.

\begin{figure*}
\centering
\includegraphics[width=.9\textwidth]{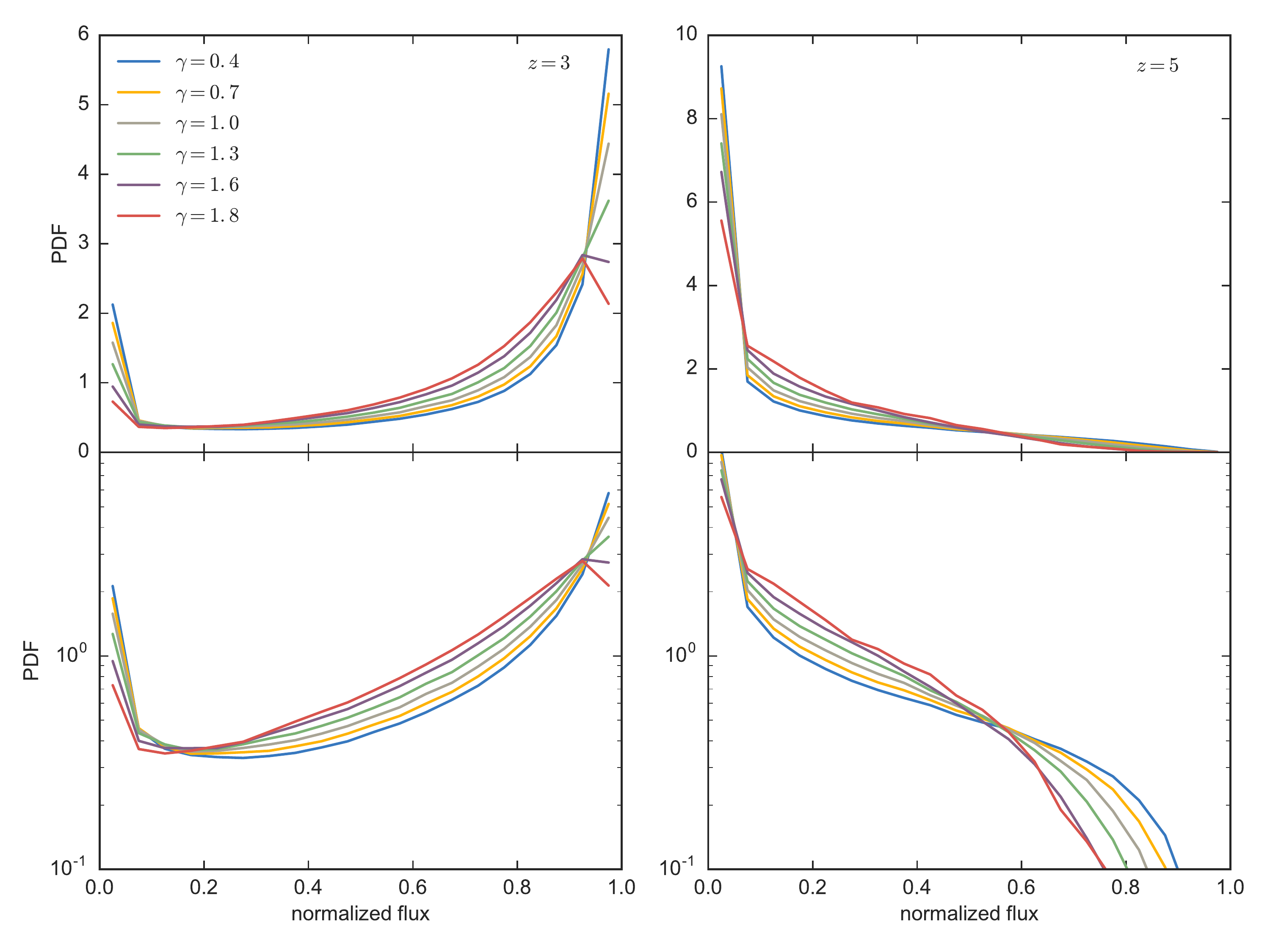}
\caption{The PDF of the transmitted flux in the \lya forest in bins of $\Delta F = 0.05$ from noisefree mock quasar spectra at two different redshifts, $z=3$ (left panels) and $z=5$ (right panels). The different colored PDF's account for different values of the slope parameter $\gamma$ of the temperature-density relation of the IGM. }\label{fig:pdf_different_gammas}
\end{figure*}

Fig.~\ref{fig:pdf_different_gammas} illustrates the effect of
changing the slope parameter $\gamma$ on the flux PDF at redshifts of $z=3$ and $z=5$ in the left and
right panel, respectively. 
Decreasing $\gamma$ increases the temperature in the underdense regions of the IGM, i.e. $\Delta<1.0$, according to the temperature-density relation (see eqn.~(\ref{eq:T-Drelation})).
Higher temperatures cause the hydrogen recombination rate to
decrease, therefore reducing the fraction of neutral \ion{H}{1} and the Ly$\alpha$ optical depth, which in
turn increases the transmitted flux in these regions. These
differences are apparent at $F > 0.5$, where the peak of the flux PDF
shifts towards higher values for lower values of $\gamma$. 
Accordingly, by measuring the \lya flux PDF one can constrain the parameter
$\gamma$.

\subsection{Empirical Quasar Continua}\label{sec:HSTdata}

In order to generate mock quasar spectra we multiply \lya absorption
fields with quasar continua. For this purpose we use the estimated
continua of $50$ real quasar spectra taken by the Hubble Space
Telescope (HST) Faint Object Spectrograph (FOS) from
\citet{Suzuki2005}, which were collected and calibrated by
\citet{Bechtold2002}. All quasars in this sample have complete
wavelength coverage from $1020$~{\AA} to $1600$~{\AA} in the rest
frame. These moderate resolution spectra ($R\sim1300$) have been
combined from several exposures, brought into the rest frame and
re-binned into pixels of size $0.5$~{\AA}.  The average signal-to-noise
ratio of the chosen quasar sample is $\langle {\rm S/N}\rangle=19.5$
per $0.5$~{\AA} pixel, and quasars with ${\rm S/N}<10$ per pixel were
removed from the sample. All quasar spectra have been renormalized to
unity at $1280$~{\AA} in the rest-frame.  The quasars in the sample
are at very low redshifts in the range $0.14 < z < 1.04$ with a mean
redshift of $\left\langle z\right\rangle=0.58$. Quasars at these low
redshifts show very few \lya absorption lines, which makes it easier
to correctly estimate their continuum level.
Spectra with broad
absorption lines or damped \lya systems were excluded from the sample,
because of the associated large uncertainties in placing their
continuum level.  In order to obtain precise and smooth continuum
estimations for these spectra, \citet{Suzuki2005} fitted Chebyshev
polynomials of different orders to each HST spectrum and applied
further fine-tuning adjustments afterwards using a $B$-spline fit. For
more details on this procedure we refer the reader to
\citet{Suzuki2005}. We use these smoothed continua as the \textit{true} continua of our mock quasar spectra. \\

\subsection{Mock Quasar Spectra}\label{sec:noise}

We multiply the empirical
quasar continua with different realizations
of the Ly$\alpha$ forest absorption field to produce mock quasar
spectra. To simplify the analysis we assume no redshift evolution of 
the quasar continua. The absorption fields are also generated at a fixed redshift, and there
is thus no redshift evolution in the \lya forest along the quasar sightline. 
These assumptions should not alter the
conclusions of this work.  Thus we generate our own mock spectra by
multiplying the \lya forest $F$ at different redshifts into the quasar
continua in the wavelength region where \lya forest is usually found,
i.e. between \lyb ($1025.18${\AA}) and \lya ($1215.67$ {\AA})
emission.

\begin{figure*}[th]
\center
\includegraphics[width=\textwidth]{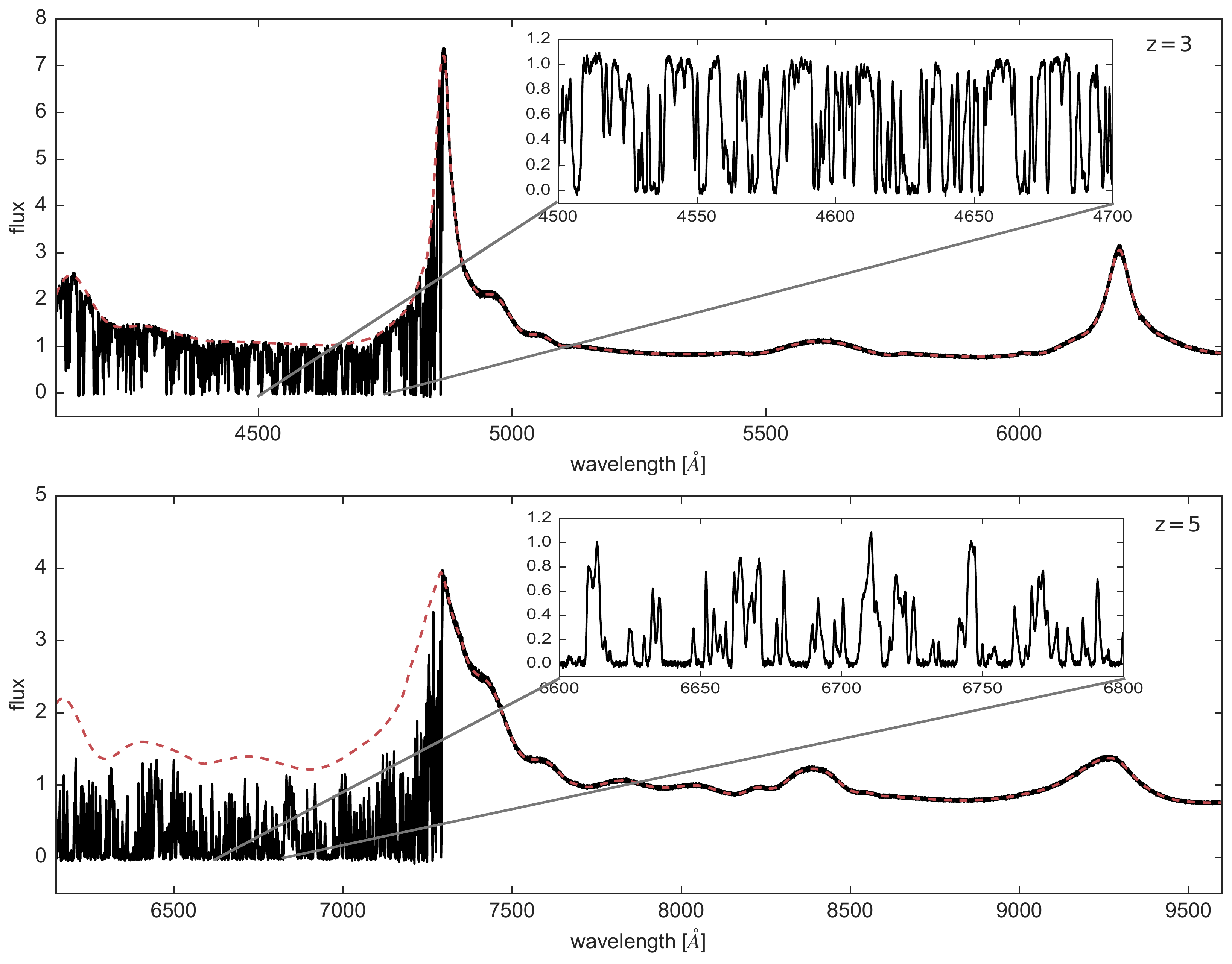}
\caption{Examples of the generated mock quasar spectra at two different redshifts, $z=3$ (upper panel) and $z=5$ (lower panel). The red dashed lines indicate the quasar continuum from one of the HST spectra from \citet{Suzuki2005}. The black curves show the resulting quasar spectra with \lya forest absorption added between the \lya and \lyb emission. The insets show a zoom into the respective \lya forests, after normalizing by the quasar continuum.
  The mock \lya forest was created assuming a value of $\gamma=1.0$. \label{fig:ex_spectra}}
\end{figure*}

We choose the pixel size and resolution of our mock spectra to be
comparable to real high resolution spectra taken with the Ultraviolet
and Visual Echelle Spectrograph (UVES) at the Very Large Telescope
(VLT) and the High Resolution Echelle Spectrometer (HIRES) from the
Keck telescope. Since the pixel size from these spectrographs is
smaller than the pixel size of $0.5$~{\AA} used by \citet{Suzuki2005},
we interpolate the HST continua onto a finer grid with the desired
pixel scale, which we chose to be $2.5$~km/s.  We then add random
white noise to our mock data, i.e. we ignore correlations between the
noise and the transmission, with a signal-to-noise ratio that is achievable with
UVES or HIRES. The highest quality quasar spectra at $z\sim 3$ from
these instruments achieve a signal-to-noise ratio in the Ly$\alpha$ 
forest of ${\rm S/N}\approx80$ per $6$~km/s resolution element \citep[see e.g.][]{Dallaglio2008, KODIAQ, OMeara2015}.
For our purposes we assume a slightly larger resolution element of $10$~km/s, which gives a
signal-to-noise ratio of 
${\rm S/N}\approx 103$ per resolution element,
assuming the same noise properties. A pixel size of $2.5$~km/s 
implies $4$ pixels per resolution element, and thus we obtain a signal
to noise ratio of $\rm S/N\approx 51$ per pixel. For quasars at $z\sim5$ we will assume a lower signal-to-noise ratio of $\rm S/N\approx 20$ per pixel, consistent with existing high resolution quasar spectra at this redshift \citep{Calverley2011}. 
However, we will see 
in \S~\ref{sec:PDF_models} that
our results at $z=5$ are fairly independent of the exact
noise level, because the \lya flux PDF at these redshifts is only
mildly sensitive to noise. We add Gaussian distributed white
noise to our mock quasar spectra with a standard deviation
$\sigma_{\rm noise}=\frac{\langle F\rangle}{\rm S/N}$, where $\langle
F\rangle$ is the mean transmitted flux and $\rm S/N\approx 51$ or $\rm S/N\approx 20$ for spectra at $z\sim 3$ and $z\sim 5$, respectively. Examples of our mock quasar spectra at redshift $z=3$ and
$z=5$ are presented in Fig.~\ref{fig:ex_spectra}.

\section{Quasar Continuum Estimation via PDF Regulation}\label{sec:pdf_regulation}

In this section we first describe the PCA basis that we will use to model the continuum of each quasar. We
then examine the impact of uncertainties in the continuum
estimation and spectral noise on the flux PDF.
Then we demonstrate the necessity of
incorporating these continuum uncertainties and noise
into models of the PDF in order to
avoid introducing biased estimates of thermal parameters. Finally, we
introduce the likelihood function $\mathcal{L_{\text{PDF}}}$ that
is the basis of our Bayesian algorithm, and show how it
enables us to simultaneously estimate the parameters governing the
flux PDF and quasar continua.

\subsection{Modeling the Quasar Continuum with Principal Component Analysis}\label{sec:pca_cont_model}

The quasar continuum is modeled using a PCA enabling us to describe
the continuum shape via a simple linear model. \citet{Suzuki2006}
analyzed the shape of $50$ HST quasar spectra and 
established a set of principal component spectra (PCS) that
reproduce the continuum shape of each quasar with high
accuracy. This accuracy in the continuum model is crucial, in order to
prevent biases when studying the statistical properties of the
Ly$\alpha$ forest.

The idea of the PCA is to represent the continuum spectrum $\ket{q_{i,
    \lambda}}$ of the $i$th quasar by a reconstructed spectrum that
consists of a mean spectrum $\ket{\mu_{\lambda}}$ and a sum of $m$
weighted PCS
$\ket{\xi_{j, \lambda}}$,
where the index $\lambda$ denotes the wavelength. Hence
\begin{align}
\ket{q_{i, \lambda}}\approx\ket{\mu_{\lambda}} + \sum_{j=1}^m \alpha_{ij}\ket{\xi_{j, \lambda}},\label{eq:pca}
\end{align}
where $\ket{\xi_{j, \lambda}}$ refers to the $j$th PCS and $\alpha_{ij}$ is its weight for quasar $i$.

\begin{figure}[t]
\center
\includegraphics[width=.5\textwidth]{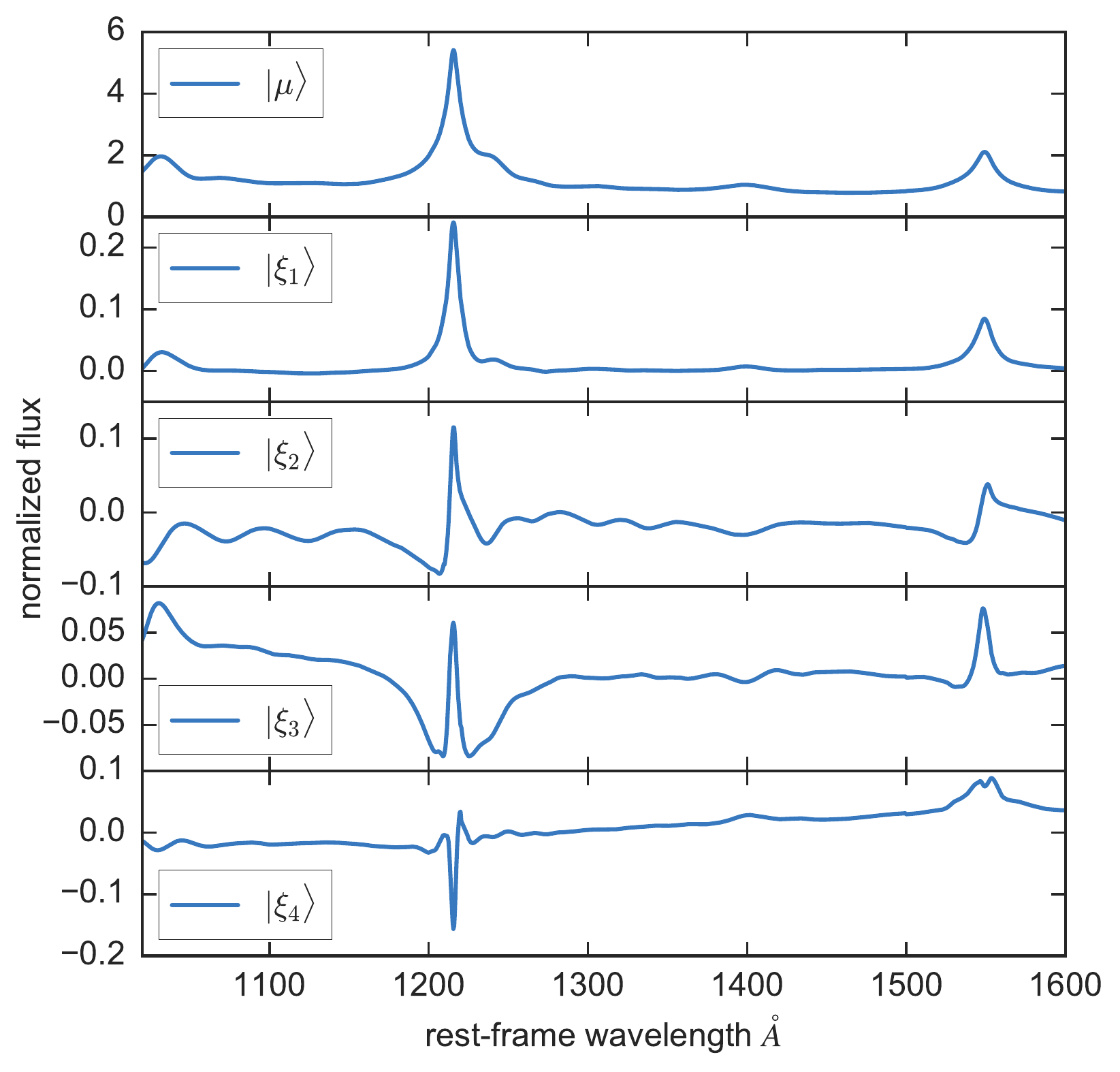}
\caption{Mean quasar spectrum $\ket{\mu}$ and the first four PCS $\ket{\xi_{i}}$ from \citet{Suzuki2006}.  \label{fig:pcs}}
\end{figure}

\citet{Suzuki2006} derived ten PCS, which are shown in Fig.~$2$
and Fig.~$3$ of their paper. We show their mean quasar spectrum and the first four PCS in Fig.~\ref{fig:pcs}. The more components are included
in the reconstruction, the lower the
variance of the individual spectra about the model fits. Table~$1$ in
\citet{Suzuki2006} summarizes the fraction of the variance accounted
for depending on the number of principal components employed.

In our analysis we will vary the number of PCS that we include when estimating the quasar continuum. Since we are only taking a finite number of components we cannot account for the total variance seen in quasar spectra and thus we inevitably introduce an error in the continuum fit. In order to analyze the best precision possible for a chosen PCA basis, we estimate the continuum model coefficients by using a  Markov Chain Monte
Carlo (MCMC) algorithm to sample the likelihood function $\mathcal{L}=\exp(-\chi^2/2)$ with 
\begin{align}
  \chi^2&= \sum_{\lambda}\frac{
    \left(C_{\text{data,} i\lambda} - C_{\text{model,} i\lambda}(\alpha_{ij})\right)^2}{\sigma_{i,\rm noise}^2}.   \label{eq:chi2}
\end{align}
Here $C_{\text{data,} i\lambda}$ is the true quasar continuum determined by \citet{Suzuki2005} (see \S~\ref{sec:HSTdata}),
$C_{\text{model,} i\lambda}$ is the model continuum from eqn.~(\ref{eq:pca}), and $\sigma_{i,\rm noise}$ is the standard
deviation of the white noise added to each spectrum.
We use the means of the posterior probability distributions of each coefficient $\alpha_{ij}$ as our best continuum model for quasar $i$. 
Note that for this estimation we take all pixels in the whole available wavelength range, i.e. between $1020$~{\AA} and $1600$~{\AA}, into account and that we do not add artificial \lya forest to the true quasar continua here. We then evaluate the precision of the estimated continuum model by calculating the continuum residuals, i.e. the relative continuum error, 
\begin{align}
\Delta C/C = \frac{C_{\text{data}, \lambda}-C_{\text{model}, \lambda}}{C_{\text{data}, \lambda}}, \label{eq:deltac}
\end{align}
only in the wavelength range that we are particularly interested in,
i.e. the \lya forest region between the \lyb and \lya emission peak. However, in order to avoid
biases due to the influence of 
proximity zones, and to guarantee that we exclude the \lyb forest in the presence of quasar redshift
errors, we only include the wavelengths between $1040$~{\AA} to $1190$~{\AA}. 
Note that all our mock spectra use the true quasar continua determined from the HST spectra.

\begin{figure*}[th]
\centering
\includegraphics[width=.75\textwidth]{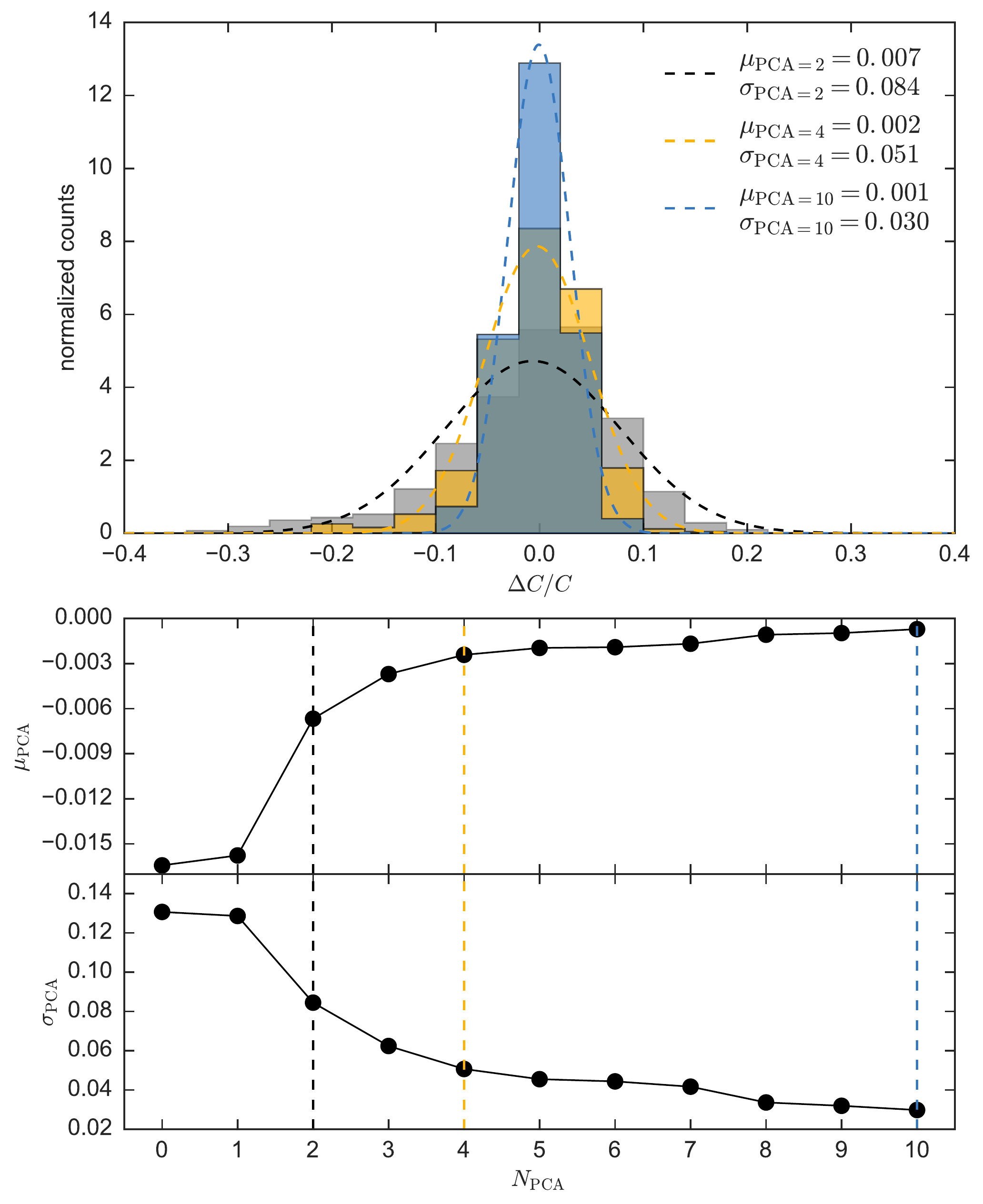}
\caption{\textit{Upper panel:} Distribution of continuum residuals, i.e. relative continuum errors, $\Delta C/C$ in the continuum estimation of $50$ HST spectra when the quasar continua are fitted to the whole available spectral range. The blue histogram shows the relative continuum errors when ten PCS where taken into account in the continuum model, whereas the yellow histogram shows the same for four components and the black histogram for two components. The dotted lines show the corresponding Gaussian distributions with the same mean and standard deviation. \textit{Middle Panel:} The mean of the continuum residual distribution versus the number of PCS, $N_{\rm PCA}$, in the continuum model. The dashed vertical lines correspond to the relative continuum errors shown in the upper panel. \textit{Bottom panel:} The standard deviation of the relative continuum errors versus the number of PCS in the continuum model. \label{fig:residuals}}
\end{figure*}

The relative continuum error $\Delta C/C$ is shown in the upper
panel of Fig.~\ref{fig:residuals} for different numbers of PCS. The distribution for the ensemble of \textit{all}
$50$ spectra follows a Gaussian distribution reasonably well. Note
that this is not necessarily the case for a fit to a single quasar spectrum. The
Gaussian distribution has a standard deviation of $\sigma_{\rm PCA}\approx
3.0\%$ when
all ten components are used to reconstruct the continuum
spectra (blue histogram). Fewer components, i.e. $N_{\rm PCA}=2$ (gray
histogram) or $N_{\rm PCA}=4$ (yellow histogram), account for less
individual variance in the spectra and thus decrease the precision of
the estimated continuum and hence increase the width of the distribution of continuum residuals.
The middle and lower panels of Fig.~\ref{fig:residuals} plot the mean
$\mu_{\rm PCA}$ and standard deviation $\sigma_{\rm PCA}$ of the relative
continuum error as a function of $N_{\rm PCA}$ used to reconstruct the continuum.
Increasing the number of PCS implies a more
precise, i.e. decreasing $\sigma_{\rm PCA}$, and less biased,
i.e. decreasing $|\mu_{\rm PCA}|$, estimation of the quasar continua
in the \lya forest region.

\subsection{Models of the PDF of the Transmitted \lya Forest Flux} \label{sec:PDF_models}

In order to calculate the PDF of the transmitted flux in the \lya
forest, the flux in the quasar spectra has to be normalized by its
continuum level. However, the exact placement of the continuum level
is very challenging, because, particularly at higher redshifts, much
of the continuum flux is absorbed by the \lya forest.
Thus one should consider the finite precision with which the continuum can be estimated.
If one instead adopts a PCA model, as is done here, this
continuum model has limited precision due to the
finite number of PCS, and thus additionally has an intrinsic uncertainty which
we quantified in the previous subsection (see Fig.~\ref{fig:residuals}).

We will now analyze the effects of a misplaced or imprecise continuum
level and the effects of noise in the spectra on the shape of the
PDF. In the presence of noise and continuum errors the continuum
normalized transmitted flux, which we denote as 
$f$, can take on values below $0$ and above $1$. We describe the level of uncertainty in the continuum normalized flux caused by the continuum error with $\sigma_C$. 
Note that we expect $\sigma_C$ to take on values comparable to or slightly
larger than the intrinsic PCA continuum model error $\sigma_{\rm PCA}$
(see Fig.~\ref{fig:residuals}), since $\sigma_{\rm PCA}$ represents the best continuum error one can achieve using the given PCA basis as the continuum model, but the continuum error $\sigma_C$ could generally be larger. 
We can approximate the transmitted flux $f$ by
modifying the perfectly normalized and noiseless flux $F$ by
adding white noise with a standard deviation $\sigma_{\rm noise}$ and
introducing the effects of the continuum errors in the following
way:
\begin{align}
f = \frac{F+\mathcal{N}(0,1)\sigma_{\rm noise}}{1+\mathcal{N}(0,1)\sigma_C},\label{eq:flux}
\end{align}
where $\mathcal{N}(0,1)$ represents a random draw from a normal Gaussian distribution with mean $\mu=0$ and standard deviation $\sigma=1$.
Note that the numerator in eqn.~(\ref{eq:flux}) adds random white noise to the perfectly normalized transmitted flux, whereas the denominator introduces the effects of continuum errors.

In Fig.~\ref{fig:residuals} we have seen that the relative continuum errors $\Delta C/C$ of the ensemble of all $50$ quasars follows
a Gaussian distribution.
Hence, we can generate PDF models that
incorporate the uncertainty in the continuum model by simply drawing Gaussian deviates
with a standard deviation of the continuum error $\sigma_C$ and
adding them to each flux pixel in the \lya forest to mock up the
continuum error. 
In this way we generate a new set of
models for the transmitted \lya forest flux PDF that incorporate the continuum errors
as well as the effects of spectral noise. 
This new set of PDF models now has 
\textit{two} free parameters: the slope parameter $\gamma$ of the
temperature-density relation of the IGM, where we consider values of $0.0\leq\gamma\leq 2.0$ in steps of $\Delta\gamma=0.01$ for our PDF models, 
and additionally, the continuum error $\sigma_C$
that we allow to vary between $0.5\%$ and $15.0\%$ in steps of
$\Delta\sigma_C=0.1\%$. The grid of PDF models is generated by averaging
over many realizations of approximately normalized mock \lya forests for every combination of
the two free parameters $\gamma$ and $\sigma_C$.

\begin{figure*}[th]
\includegraphics[width=\textwidth]{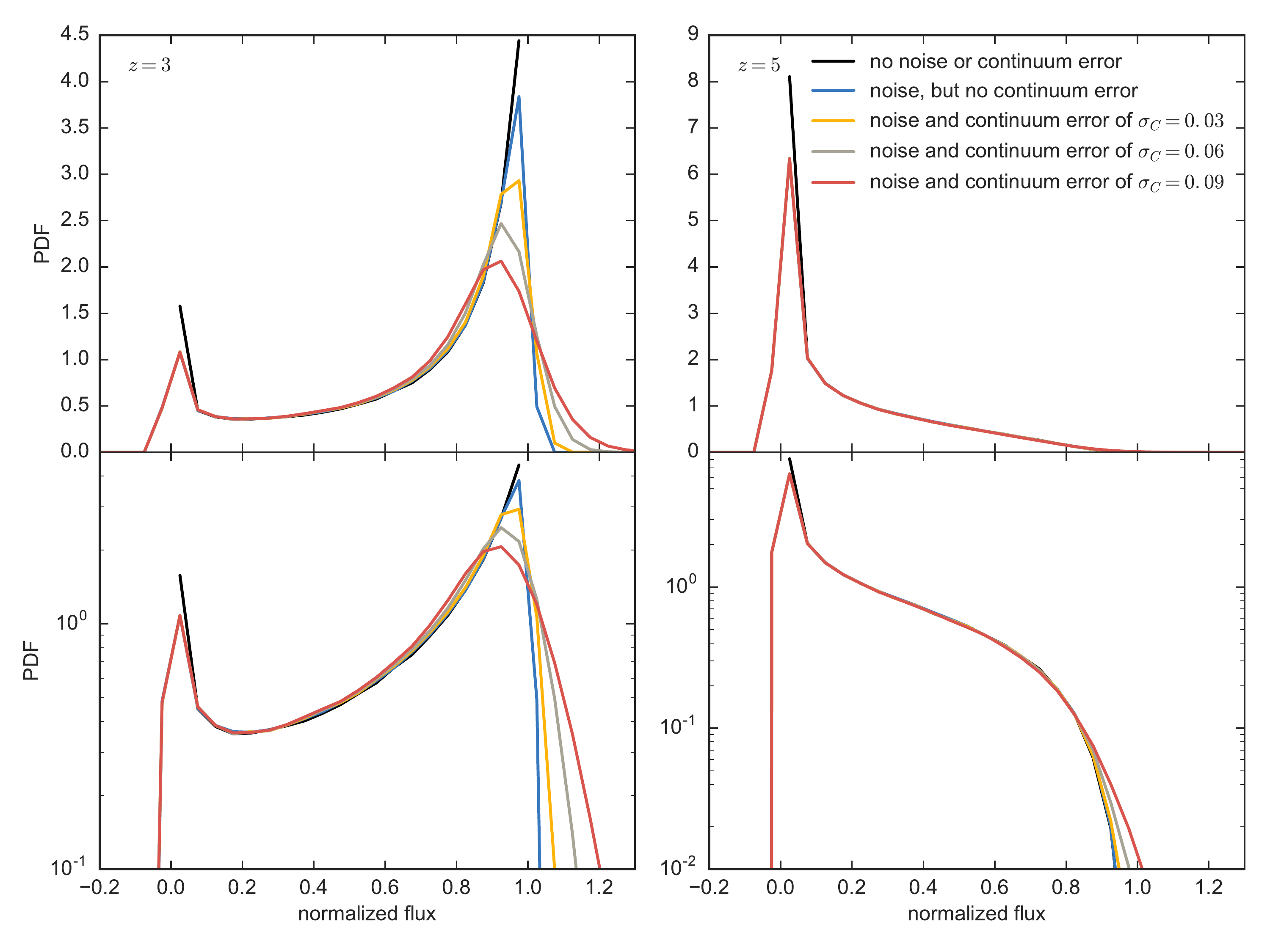}
\caption{PDF's of the flux in the \lya forest at redshifts of $z=3$ (left panels) and $z=5$ (right panels) with different levels of continuum error and noise added to the quasar spectra for a fixed value of $\gamma=1.0$. The black curves show the unrealistic case of the PDF with noiseless spectra and perfectly known continua. The blue curves show the PDF when random white noise was added to the quasar spectra, but we still assume perfect knowledge of the continuum level. The yellow, gray and red curves show the PDF's with different levels of continuum error, i.e. $\sigma_C=3\%$, $\sigma_C=6\%$ and $\sigma_C=9\%$, and noise added to the spectra, respectively. The plot shows clearly that at a redshift of $z=3$ the effects of noise and continuum error in the continuum estimation change the shape of the PDF, hence by fitting for the PDF one can obtain constraints on the continuum error. At higher redshifts of $z=5$ the effects of noise and continuum error do only mildly alter the shape of PDF, since the differences in the blue, yellow, gray and red curves are very small. \label{fig:PDF_cont_error}}
\end{figure*}

In Fig.~\ref{fig:PDF_cont_error} we demonstrate the effects of
continuum error and noise on the flux PDF at redshifts $z=3$
(left panels) and $z=5$ (right panels), for a
fixed value of $\gamma=1.0$. The different colored curves
represent PDF models constructed from mock \lya forest 
spectra with different levels of continuum error added. The black curve shows the
PDF taken from perfect mock \lya forests without any continuum error or noise.
There is an obvious difference between the PDF's at redshift $z=3$
and those at $z=5$. Whereas at redshift
$z=3$, one can clearly distinguish the different PDF models
depending on the underlying continuum error, these differences are
hardly noticeable at redshift $z=5$. The continuum errors have only a weak
impact on the shape of the PDF at the higher redshift, which
can be understood as follows.

The PDF is a function of the approximately normalized \lya forest flux
$f$, 
which itself is a function of the continuum error as shown
in eqn.~(\ref{eq:flux}). Thus:
\begin{align}
\Delta\text{PDF}=\frac{\diff\text{PDF}}{\diff f}\frac{\diff f}{\diff \sigma_C} \Delta \sigma_C\label{eq:diff}
\end{align} 
The first term
$\frac{\diff\text{PDF}}{\diff f}\approx \frac{\diff\text{PDF}}{\diff F}$
indicates that the change in the PDF
is dependent on the slope of the PDF itself, thus the steeper the PDF
is, the larger the change with varying $\sigma_C$. This can be clearly
seen in the PDF at $z=3$ (left panel of Fig.~\ref{fig:PDF_cont_error}): at fluxes around $0.8\lesssim f\lesssim 1.0$ the PDF has a steep slope such that the
term $\frac{\diff\text{PDF}}{\diff f}$ is large, as are the differences in the PDF models
for the different values of $\sigma_C$. Whereas at 
smaller flux values of $0.2 \lesssim f \lesssim 0.8$ the slope
of the PDF $\frac{\diff\text{PDF}}{\diff f}$ is smaller,
as are the differences in the PDF models. But that
explanation is not sufficient, because the PDF is also steep for very
small flux values around $f\approx0.1$ at $z=3$ as well as at
$z=5$. At these small fluxes there is no change in the PDF models visible for
different values of $\sigma_C$ because of the second term of
eqn.~(\ref{eq:diff}) $\frac{\diff f}{\diff \sigma_C}=\left|\frac{f}{1+\mathcal{N}(0,1)\sigma_C}\right|$ that indicates
the change in the PDF is also dependent on the actual flux
level $f$.
Hence for small fluxes $f\approx0.1$ there is a much smaller
change in the PDF than at larger flux values $f\approx1.0$. Thus at $z=5$ where the flux
values are lower overall (the mean flux is
$\langle F \rangle = 0.189$), 
there is
very little change in the PDF when the continuum error $\sigma_C$
is increased. This implies that at lower redshifts the PDF is very sensitive
to continuum errors, whereas at higher redshift the sensitivity is much lower,
which is illustrated in Fig.~\ref{fig:PDF_cont_error}.

\subsection{The Likelihood Function}\label{sec:lpdf}

In order to estimate model parameters from measurements of the flux PDF, we introduce 
a likelihood function $\mathcal{L_{\text{PDF}}}$. The PDF
models are explicitly dependent on two free parameters, the physical
parameter $\gamma$ and the continuum error 
$\sigma_C$. In addition there is an implicit
dependence on the parameters of the quasar continuum model, i.e. the
coefficients $\alpha_{ij}$ of the PCS for each quasar
(see eqn.~(\ref{eq:pca})), since each spectrum needs to be divided by
its estimated continuum level in order to normalize the transmitted
\lya forest flux before calculating the flux PDF.  Thus the likelihood
function has a multi-variate Gaussian form and is given by the
following 
equation:
\begin{align}
 \mathcal{L_{\text{PDF}}} &= \mathcal{L}\left(\gamma, \sigma_C, \alpha_{ij}|\text{\textbf{PDF}}_{\text{data}}\right)\nonumber\\
 &=\frac{1}{\sqrt{\det(\textbf{C})}}\exp\left(-\frac{1}{2}\textbf{d}^{\rm T} \textbf{C}^{-1}\textbf{d}\right)\label{eq:L_PDF}.  
\end{align}
Here $\textbf{C}$ represents the covariance matrix governing the covariances between the different flux bins in the PDF. The
vector $\textbf{d}$
denotes, 
the difference between the calculated flux PDF and the model PDF for each bin:
\begin{align}
\textbf{d} = \text{\textbf{PDF}}_{\text{data}}-\text{\textbf{PDF}}_{\text{model}}(\gamma, \sigma_C, \alpha_{ij}). 
\end{align}

Our new Bayesian formalism is based on this likelihood function
$\mathcal{L_{\text{PDF}}}$ which we sample with a MCMC algorithm. The goal of an MCMC is to draw samples from a
complex probability distribution for which direct sampling is
challenging, thus providing an approximate representation of the true
posterior probability distribution of the model parameters. For our
MCMC algorithm we employ the python implementation \textit{emcee},
which is described in detail by \citet{emcee}. An important advantage
of this algorithm compared to other codes is its affine-invariance
property, first proposed by \citet{Goodman2010}. This makes the
algorithm insensitive to any covariances among the parameters and
requires hand-tuning of only one or two parameters compared to $\sim
N^2$ for a traditional algorithm in an $N$-dimensional parameter
space.  Another useful property of the \textit{emcee} implementation
is that the parameter space is explored by a set of chains, so-called
walkers, that evolve in parallel as an ensemble, rather than by a
single chain. At every step each walker is randomly assigned to a
partner walker, moving along lines which connect the single chains to
each other. This feature makes the exploration of the parameter space
extremely efficient and allows for parallel computing on multi-core
processors.

\subsection{The Covariance Matrix $\textbf{C}$}\label{sec:covariance}

There are various sources of covariance in the flux PDF. 
The flux in the \lya forest contains spatial correlations that give rise to
covariances in the flux PDF. The continuum residuals due to continuum estimation errors of the quasar continua will also result in
non-diagonal entries in the covariance matrix of the flux PDF. In what follows we will
investigate a set of models for the quasar continua that
take different numbers of PCS into account, i.e. we vary $N_{\rm
  PCA}=0, 1, 2, 3, 4$. For each of the $N_{\rm PCA}$ that we explore
we generate a separate covariance matrix.

To generate the covariance matrix we create a large set of mock quasar
spectra as described in \S~\ref{sec:mock_spectra}. For each mock
spectrum, the underlying continuum, which is a draw from the 50 HST spectra, is
fit with $N_{\rm PCA}$ PCS in the absence of \lya forest absorption
(see \S~\ref{sec:pca_cont_model}). The mock spectrum, which is
a product of the true HST
continuum times a realization of the \lya forest, is divided by the approximate PCA
continuum model, thus giving approximately normalized \lya forest spectra
with a continuum error that is intrinsic to the continuum model
itself. We create many ensembles of $N_{\rm QSO}$ quasar spectra generated in this way, 
and calculate the PDF of each realization. We use this ensemble of PDF's to
calculate the covariances between the different PDF
flux bins and obtain a covariance matrix. This covariance matrix now
contains the continuum error that is intrinsic to the
continuum model with the chosen number of PCS. We chose a fixed value
for $\gamma=1.0$, which we set to be our fiducial value for the mock
data set, when generating the covariance matrices.

Note that we consider bins in the PDF in the flux range of
$-0.5\leq f\leq1.5$. In the absence of continuum error and noise only
the bins between $0.0 \leq f \leq 1.0$ would be populated. The inclusion of
continuum error and noise will thus populate bins below zero and
above one, but for any given $\gamma$ and $N_{\rm PCA}$ not all flux
bins between $-0.5\leq f\leq1.5$ will be populated.  Indeed the very low and
very high end of this range rarely contain any flux pixels, and for a
finite number of samples these bins may not be populated such that the
resulting covariance matrix is zero. However, models with values of
$\gamma$ significantly different than those used to
generate the covariance matrix may actually populate these bins, such
that we cannot just neglect the bins and reduce the allowed flux
range.  The correct approach would clearly be to adopt a more
complicated procedure and compute a different covariance matrix for
each model, rather than simply fixing $\gamma$ and $N_{\rm
  PCA}$. However, we lack the ability to generate covariance matrices
for models with different values of $\sigma_C$, other than those
intrinsic to the continuum model, i.e. $\sigma_{\rm PCA}$, 
since by construction adding a random level of
continuum error as described in eqn.~(\ref{eq:flux}) has no
covariance. In this way our covariance matrix is approximate, because
we adopt a model with fixed $N_{\rm PCA}$ and an implied single value of
$\sigma_C = \sigma_{\rm PCA}$
to calculate it. This means that the range of values of
$\sigma_C$ we can explore to create covariance matrices is limited and
thus results in matrix entries with zero covariance at the edges of
the considered flux range.

\begin{figure*}[th]
\centering
\includegraphics[width=1.\textwidth]{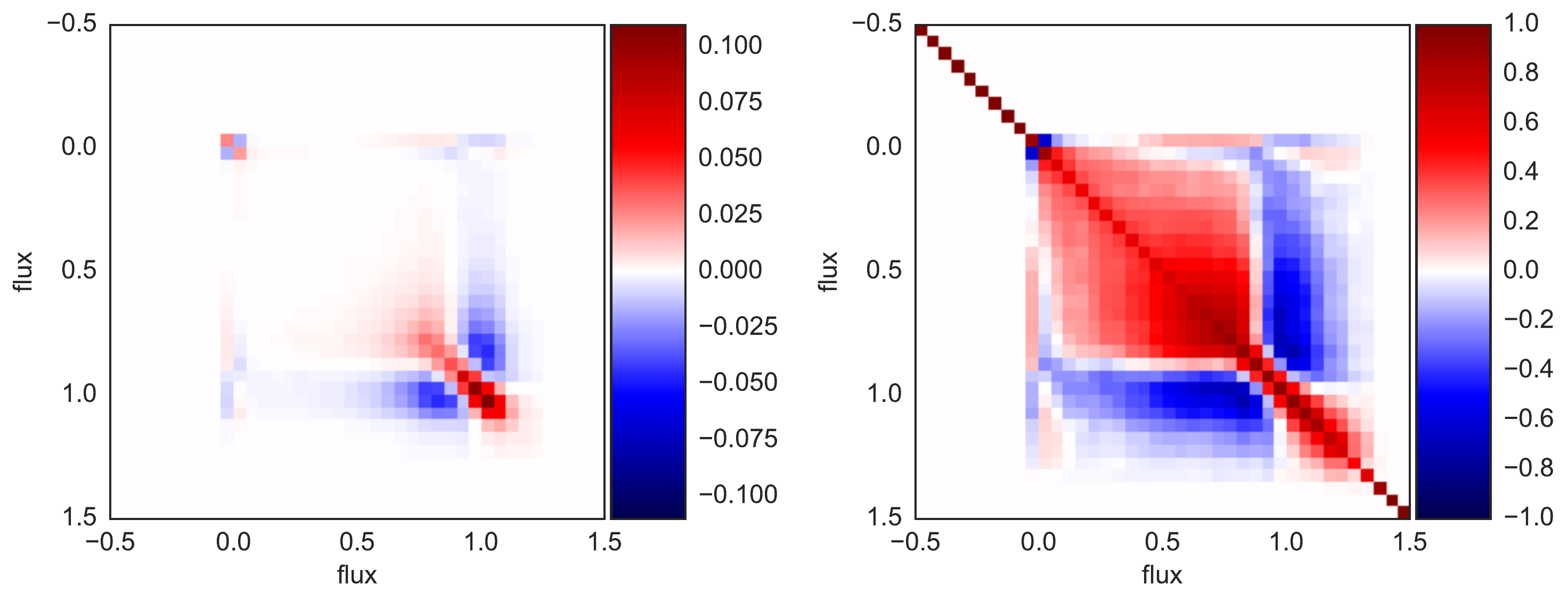}
\caption{Covariance and correlation matrix shown in the left and right panel, respectively. The matrices are created from the PDF's of ten mock quasar spectra, i.e. $N_{\rm QSO}=10$, at redshift $z=3$ with a fiducial value of $\gamma=1.0$. Two PCS, i.e. $N_{\rm PCA}=2$, are included in the continuum model for normalizing the quasar spectra before calculating the PDF, i.e. the matrices incorporate a continuum error of $\sigma_{\rm PCA=2}\approx8.4\%$. \label{fig:covariance}}
\end{figure*}

For simplicity, we opt to proceed with a single covariance matrix (for each employed value of
$N_{\rm PCA}$). In order to prevent the covariance matrix from becoming
singular, we apply a technique known as matrix shrinkage and add
an identity matrix multiplied with a small
value \citep[e.g.][]{matrix_shrinkage} to our covariance, thus setting
a floor on the diagonal elements.
The size of the floor should be roughly the same size as the
smallest calculated covariance and after some experimentation, we set this value
to $10^{-4}\times(N_{\rm QSO}/10)$, where $N_{\rm QSO}$ is the number of quasars in the data set that is analyzed. 
We will come back to this subtle point 
in \S~\ref{sec:results} when discussing the effects of this floor on our results.
Fig.~\ref{fig:covariance} shows the covariance and correlation matrices of the PDF for an ensemble
of ten quasar spectra at redshift $z=3$ generated for a continuum model containing two PCS.

\subsection{Effects on the Estimation of $\gamma$ Due to Continuum Errors and Spectral Noise}

\begin{figure*}[t]
\centering
\includegraphics[width=.9\textwidth]{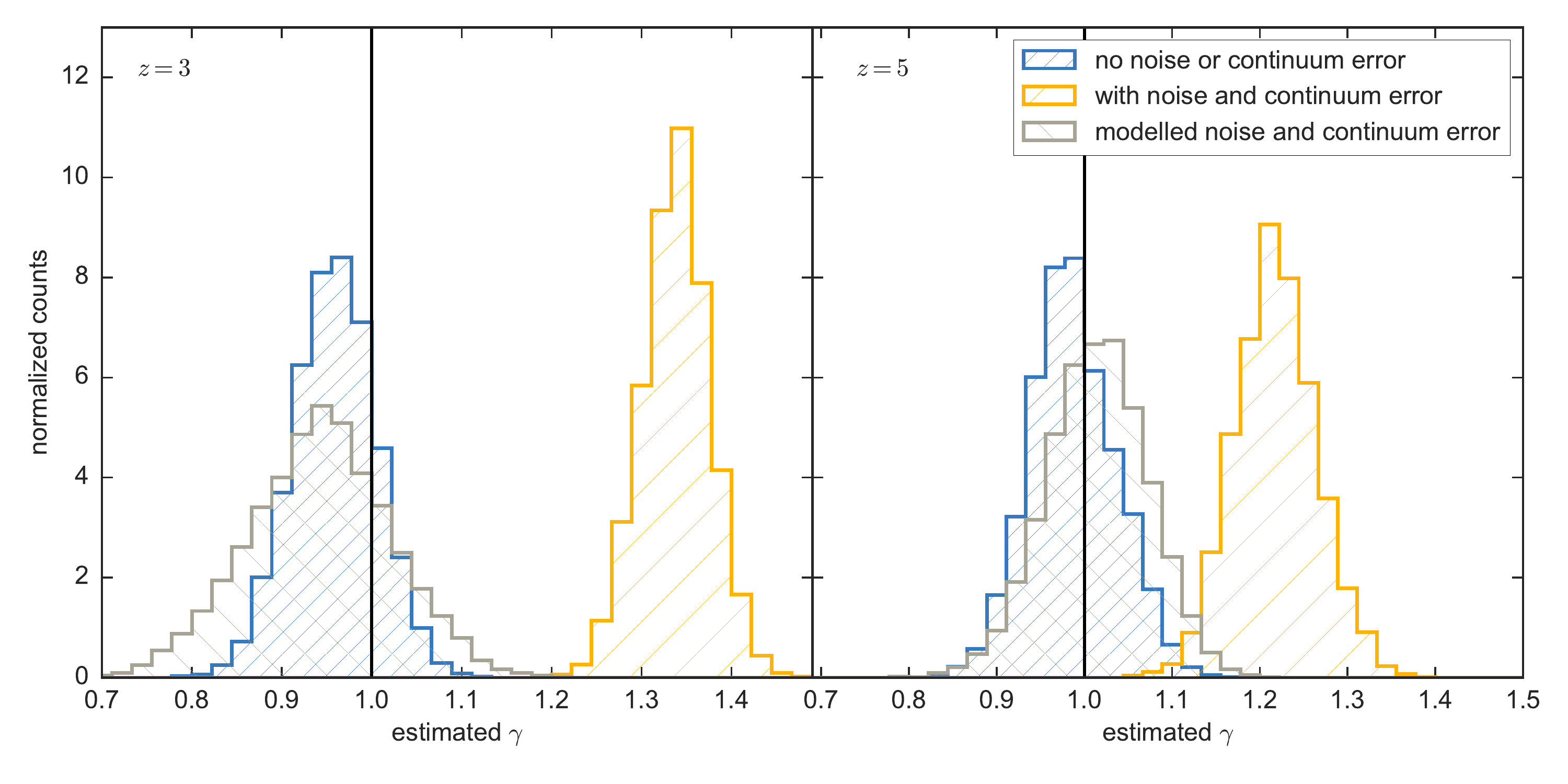}
\caption{Posterior probability distribution of $\gamma$ for ten realisations of the \lya forest at redshift $z=3$ (left panel) and $z=5$ (right panel). The black vertical line indicates a value of $\gamma=1.0$, which was used as our fiducial value to generate the quasar data. The blue histogram indicates the posterior probability distribution of $\gamma$ when assuming a perfectly known continua and noise-free spectra. Introducing continuum errors and spectral noise to the \lya forest spectra results in a strongly biased distribution (yellow). The gray histogram shows that this bias can be resolved again by modelling the continuum errors and spectral noise into our PDF models and covariance matrices and treating the continuum error $\sigma_C$ as an additional free parameter that can be estimated and marginalized out. \label{fig:bias}} 
\end{figure*}

\begin{figure*}[t]
\centering
\includegraphics[width=.8\textwidth]{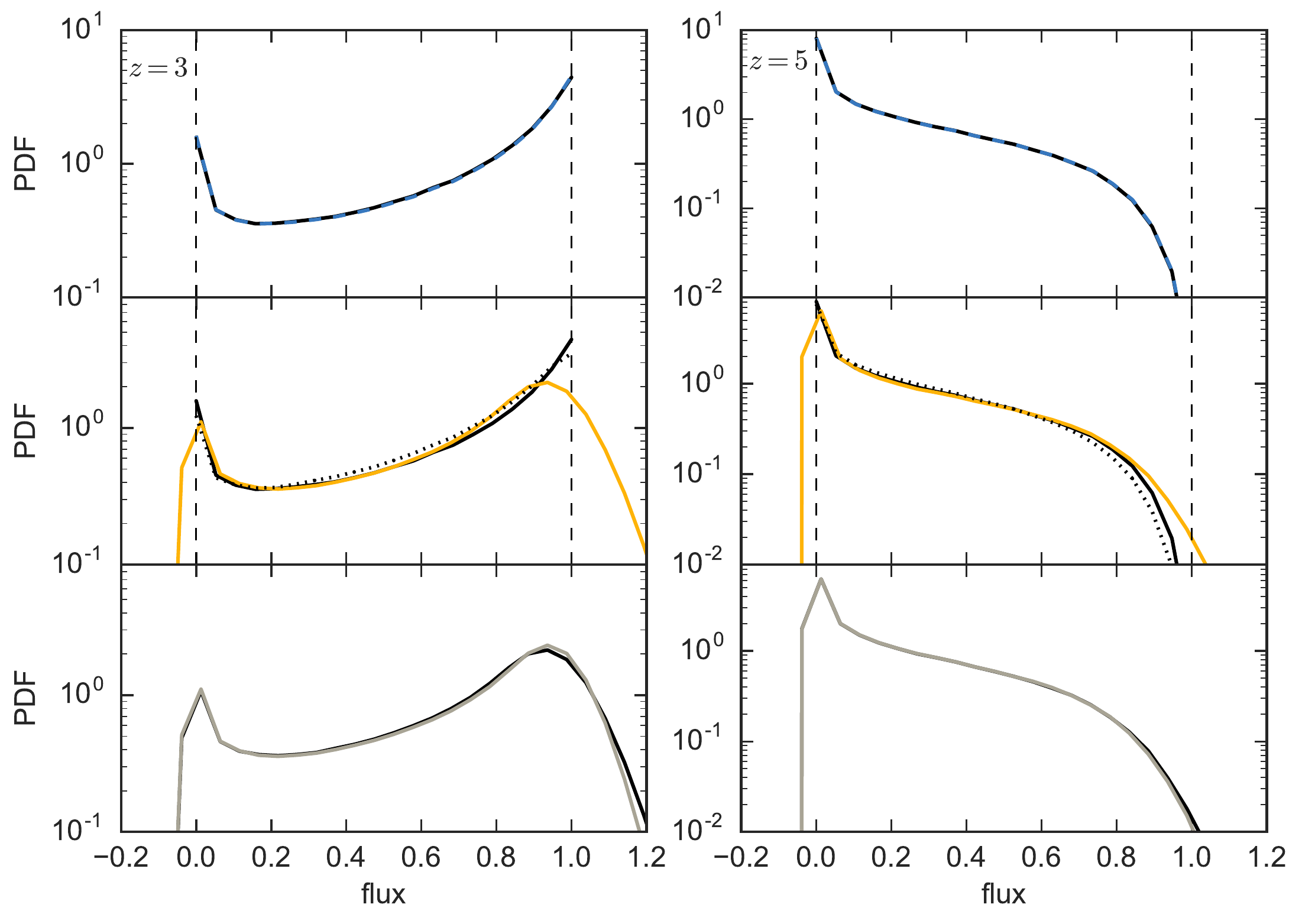}
\caption{Ly$\alpha$ flux PDF's corresponding to the three described scenarios from Fig.~\ref{fig:bias} at $z=3$ (left panels) and $z=5$ (right panels). \textit{Upper panels:} For noise-free quasar spectra with perfcetly known continua our algorithm recovers the true PDF model (black curves) very well, i.e. the recovered PDF models shown as the blue dashed curves and black curves agree. \textit{Middle panels:} Adding noise and the effects of continuum error to the quasar spectra leads to a discrepancy between the true PDF models (black curves) and the recovered PDF models (yellow curves) that rather agree with PDF models that corresponds to a biased value of $\gamma$ (black dotted curves). The black dashed lines show the considered flux range for estimating the best PDF model. \textit{Bottom panels:} When augmenting the considered flux range in the PDF models and the covariance matrices the recovered PDF models (gray curves) agree again with the true PDF models (black curves) and the fiducial value of $\gamma$ can be recovered. \label{fig:bias_KG}} 
\end{figure*}

We have shown previously that we will have to take continuum errors into account when analyzing the flux PDF's of the normalized quasar spectra. We have also
shown how we modify the covariance matrix (\S~\ref{sec:covariance}) and PDF models (\S~\ref{sec:PDF_models}) to include the continuum
error due to the intrinsic limitations of the continuum model. We will
now demonstrate that these steps are necessary to obtain unbiased
estimates of the thermal parameter $\gamma$ that is our main
interest. Note that --- in this subsection only ---
we do not use our full PDF regulated continuum fitting
algorithm for illustrating the importance of incorporating
the continuum errors, but for the
sake of simplicity we conduct an exercise with a simplified version of
our method that only estimates the parameter $\gamma$ and $\sigma_C$,
but no continuum parameters. Rather we analyze already normalized quasar spectra that have been divided by their best fit continuum model with $N_{\rm PCA}=2$ that was estimated when no \lya forest absorption was added to the spectra (see \S~\ref{sec:pca_cont_model}), thus resulting in approximately normalized \lya forest spectra.

Our MCMC based algorithm produces posterior probability distributions
for the free parameters that we would like to estimate, in this
particular case only $\gamma$ and $\sigma_C$. 
We will now examine three different scenarios that we will explain in the following paragraph. 

Fig.~\ref{fig:bias} shows the resulting posterior probability distribution of the
parameter $\gamma$ when running our simplified algorithm for the three
different situations for an ensemble of ten quasar spectra. The right
and left panel of the figure correspond to the analysis at redshift
$z=3$ and $z=5$, respectively. The histogram shown in blue represents
the first situation which is the ideal case, where there is neither
spectral noise nor continuum error, i.e. $\sigma_C=0$, included in the \lya forest spectra. 
Also the PDF models for this
ideal case and the covariance matrix assume noise-free spectra without continuum error and thus only contain entries between
$0\leq F \leq 1$. The covariance matrix used in this case was calculated from multiple PDF's assuming noise-free, perfectly continuum normalized spectra. In this ideal situation our algorithm produces an
unbiased estimate for the parameter $\gamma$ for both redshifts and
recovers the fiducial input value of $\gamma=1.0$ shown as the black
vertical line.  The posterior probability distribution is not exactly
centered around the fiducial value which is due to a random
fluctuations in the realization of a finite number of quasar spectra.

Of course this first scenario is very unrealistic, because real quasar
spectra contain noise and we do not know their continuum level precisely and hence do not have perfectly normalized \lya forests. 
Thus in the second scenario, shown as the yellow histogram, spectral noise and
continuum error with $\sigma_C=\sigma_{\rm PCA=2}\approx 8.4\%$ were added to the \lya forest spectra by dividing each quasar spectrum by its continuum model with $N_{\rm PCA}=2$. 
However, in the model fitting, we still use the PDF model and the covariance matrix corresponding
to the ideal case of noise-free spectra with no continuum error. 
Thus the PDF of the approximately normalized \lya forest spectra now contains flux pixels below $f<0$ and above $f>1$, but then all pixels outside the flux
interval $f\in[0, 1]$ are neglected in the likelihood function.
Fig.~\ref{fig:bias} shows that neglecting the noise and continuum error in the modeling clearly
results in a very biased estimate for the parameter $\gamma$, with the
true fiducial value of $\gamma=1.0$ lying far outside the resulting
posterior probability distribution (yellow histogram).

This example demonstrates the need to incorporate the
effects of spectral noise and continuum error into the PDF models and the
covariance matrix. Hence we enlarge the range of fluxes that we consider 
and add these effects as described in \S~\ref{sec:PDF_models} and \S~\ref{sec:covariance} to the models for the
PDF and the covariance matrices (for $N_{\rm PCA}=2$, i.e. $\sigma_C=\sigma_{\rm PCA=2}\approx 8.4\%$). We again estimate the posterior
distribution for $\gamma$, now marginalizing over $\sigma_C$ as a
nuisance parameter. The \lya forest spectra still contain noise and continuum
errors but these effects are now included in the models of the
flux PDF and the covariance matrix. The posterior probability distribution for $\gamma$ in this third scenario is shown in
Fig.~\ref{fig:bias} as the gray histogram. The bias in the $\gamma$ estimate that arose in the previous
case, is now removed, allowing us to
recover the fiducial input value, albeit with slightly lower precision than in the
idealized first case.

Incorporating continuum error into our PDF models introduced the
free parameter $\sigma_C$ into our formalism. Thus the posterior
probability distribution for $\gamma$ in this last case is
marginalized over this parameter. This marginalization and the reduced
precision due to spectral noise causes the posterior
probability distribution to broaden, which can be seen when comparing the
blue and gray histograms. The resulting reduction in precision is greater
at $z=3$, but less obvious
at redshift $z=5$, because continuum errors have a much smaller
impact on the PDF models at higher redshift (see
Fig.~\ref{fig:PDF_cont_error}). Indeed, the bias that arose in the second scenario when
adopting the idealized (noiseless spectra and perfect continuum) models and covariance matrix at redshift
$z=5$, were largely due to the effects of ignoring the noise in modeling, rather than
the continuum error.

Fig.~\ref{fig:bias_KG} shows the PDF models corresponding to the three different scenarios described in Fig.~\ref{fig:bias}. The black curves in each panel show the true PDF that ideally should have been recovered by our algorithm, whereas the colored curves show the actual recovered PDF. In the first and third case the recovered PDF models (blue dashed and gray curves) agree with the true PDF models (black curves)very well and lead to an unbiased estimate of the parameter $\gamma$ (see Fig.~\ref{fig:bias}). The middle panels show the second scenario, where the quasar spectra contain noise and continuum error and thus the resulting PDF's (yellow curves) contain flux values outside of the considered flux range of $0\leq f\leq 1$, that is indicated by the black dashed lines. The resulting PDF models differ from the true PDF models (black curves) and prefer PDF models that correspond to biased values for $\gamma$ (dotted black curves). Considering a wider range of flux values and modeling the effects of noise and continuum error in the PDF's and covariance matrices (bottom panels) resolves this discrepancy between the true and the recovered PDF models and the fiducial value of $\gamma$ can be recovered again. 

We now have the tools to obtain unbiased estimates for thermal
parameters of the IGM such as $\gamma$ and can also estimate the
uncertainty $\sigma_C$ inherent to imperfect continuum fits. In the
following section we will apply our full method to ensembles of mock quasar
spectra.

\section{Results}\label{sec:results}

In this section we evaluate the efficacy of our new Bayesian formalism
for simultaneously estimating the continuum models of an ensemble of quasar
spectra, the continuum error, and the thermal properties of the IGM at two different
redshifts, $z=3$ and $z=5$.

\begin{figure}[h]
\centering
\includegraphics[width=.5\textwidth]{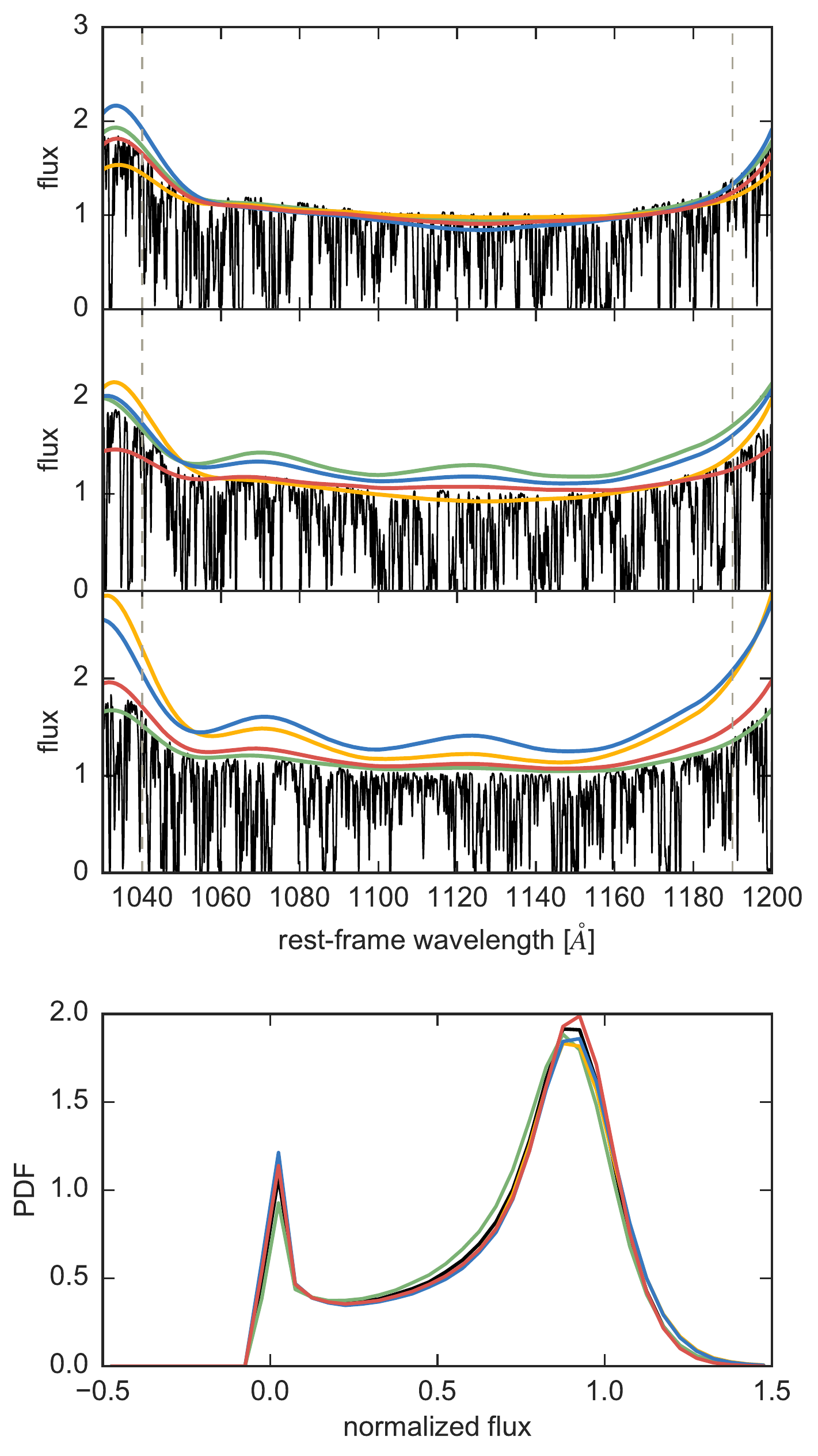}
\caption{The three upper panels show different quasar spectra from the ensemble in black (for better visibility we show the noise-free spectra). The different colored lines show different estimates for their continuum models, estimated in the region between the two gray dashed lines, that indicate the \lya forest region that we take into account, excluding the effects of proximity zones. The coefficients of the continuum model are random draws from their posterior probability distributions, the different colors representing the same draw. The lower panel shows the corresponding PDF's. The black curve demonstrates the true PDF when assuming the fiducial input parameters. The colored curves show the resulting flux PDF's of the normalized quasar spectra (and the seven other spectra of the ensemble), the colors corresponding to the same random draw from the posterior porbability distributions. }\label{fig:variation}
\end{figure}

The implicit dependence of the likelihood $\mathcal{L_{\text{PDF}}}$
in eqn.~(\ref{eq:L_PDF}) on the continuum model parameters
$\alpha_{ij}$ arises because each quasar spectrum is divided by its
continuum model before the \lya forest flux PDF can be calculated.
The true underlying flux PDF should be the same for an ensemble of
quasar spectra at the same redshift, and independent of the continuum
parameters. By forcing these quasar spectra to result in the same flux
PDF their continua will be regulated by the PDF estimation itself.  If
one quasar continuum is significantly over- or underestimated the
resulting PDF would contain too many low or high flux pixels and the
likelihood function $\mathcal{L_{\text{PDF}}}$ would disfavor the
corresponding PDF model. In this way we have a method to
simultaneously estimate the unknown continuum parameters $\alpha_{ij}$
of each quasar, the thermal parameters governing the flux PDF (for our
simplified model this is presently only the parameter $\gamma$), and
the underlying continuum error $\sigma_C$. The parameter space thus
quickly becomes very large, since for $N_{\rm QSO}$ quasar spectra,
each quasar continuum is fit with $N_{\rm PCA}$ principal components,
resulting in a $(N_{\rm QSO}\times N_{\rm PCA}+2)$-dimensional
parameter space.  Fig.~\ref{fig:variation} illustrates the basic idea
of our PDF regulated continuum fitting approach. The upper panels show
three quasar spectra in black from an ensemble of ten quasar spectra and 
different estimates of their continua, which are random draws from the posterior probability distributions of the $N_{\rm PCA}=2$ continuum model coefficients, where the colors of the curves all indicate the same draw. Note that the continuum models are only fit to the data in
between the two gray dashed lines in the region of the \lya forest
between $1040${\AA} and $1190${\AA}. The posterior probability distribution of the continuum model coefficients are obtained via our MCMC based algorithm. 
The lower panel depicts the resulting PDF's when
normalizing the quasar spectra by their respective continuum models. The
black PDF represents the true PDF when assuming the fiducial input parameter of $\gamma=1.0$ and $\sigma_C=\sigma_{\rm PCA=2}=8.4\%$. The colored PDF's are from the same random draw as the continuum models of the quasar spectra. 
Note that the PDF's include the \lya forest flux pixels from
the whole ensemble of ten quasar spectra and not only from the three example
spectra shown in the upper panels.

Note that in this paper we consider three different errors. The continuum error that is intrinsic to the continuum model is called $\sigma_{\rm PCA}$ and is due to the finite number of PCS. The estimated continuum error $\sigma_C$ is the quantity we estimate with MCMC and want to marginalize out in the end. It is mainly due to the finite precision of the PCS plus it reflects also the inability to precisely determine the continuum in the presence of \lya absorption. The third error we are considering is the width of the distribution of continuum residuals $\Delta C/C$ (see eqn.~(\ref{eq:deltac})), i.e. the actual PDF regulated continuum error $\sigma_{\Delta C/C}$. 
This gives an estimate of the error between the true quasar continuum and the estimated continuum model. It can only be calculated when dealing with mock data since we need to know the true quasar continuum. Thus we can test later on whether our estimated continuum error $\sigma_C$ reflects the actual PDF regulated continuum error $\sigma_{\Delta C/C}$. 

First we will investigate how well our algorithm performs on a single ensemble of ten high-resolution quasar spectra, and
explore the degeneracies between the parameters that govern the shape of the PDF. We conclude with an assessment of
the reliability of our method by applying it to 100 different realizations of ensembles of ten quasar spectra.

\subsection{Analysis of an Ensemble of Quasar Spectra}\label{sec:res_single}
 
We investigate the efficacy of our algorithm for an ensemble of ten
high-resolution quasar spectra, i.e. $N_{\rm QSO}=10$, which is a
realistic ensemble size corresponding to the path length of a typical
redshift bin given current high-resolution spectroscopic quasar
samples \citep{Waltherinprep}. The continuum models that we analyze
have different numbers of PCS, namely $N_{\rm PCA}=0, 1, 2, 3, 4$. No
principal components $N_{\rm PCA}=0$, means that we divide each quasar
by the mean quasar spectrum from \citet{Suzuki2006} and only fit for
the PDF parameters $\gamma$ and $\sigma_C$. Thus the largest parameter
space we are considering corresponds to $N_{\rm QSO}=10$ quasar
spectra with each continuum modeled with $N_{\rm PCA}=4$, which
results in a $42$-dimensional parameter space. Due to the long
computational time that it takes until convergence is achieved, we did
not consider continuum models with more components. Also,
Fig.~\ref{fig:residuals} shows that we do not expect a big improvement
in the precision of the continuum estimation for continuum models with
more than $4$ PCS.

\begin{figure*}
\centering
\includegraphics[width=.8\textwidth]{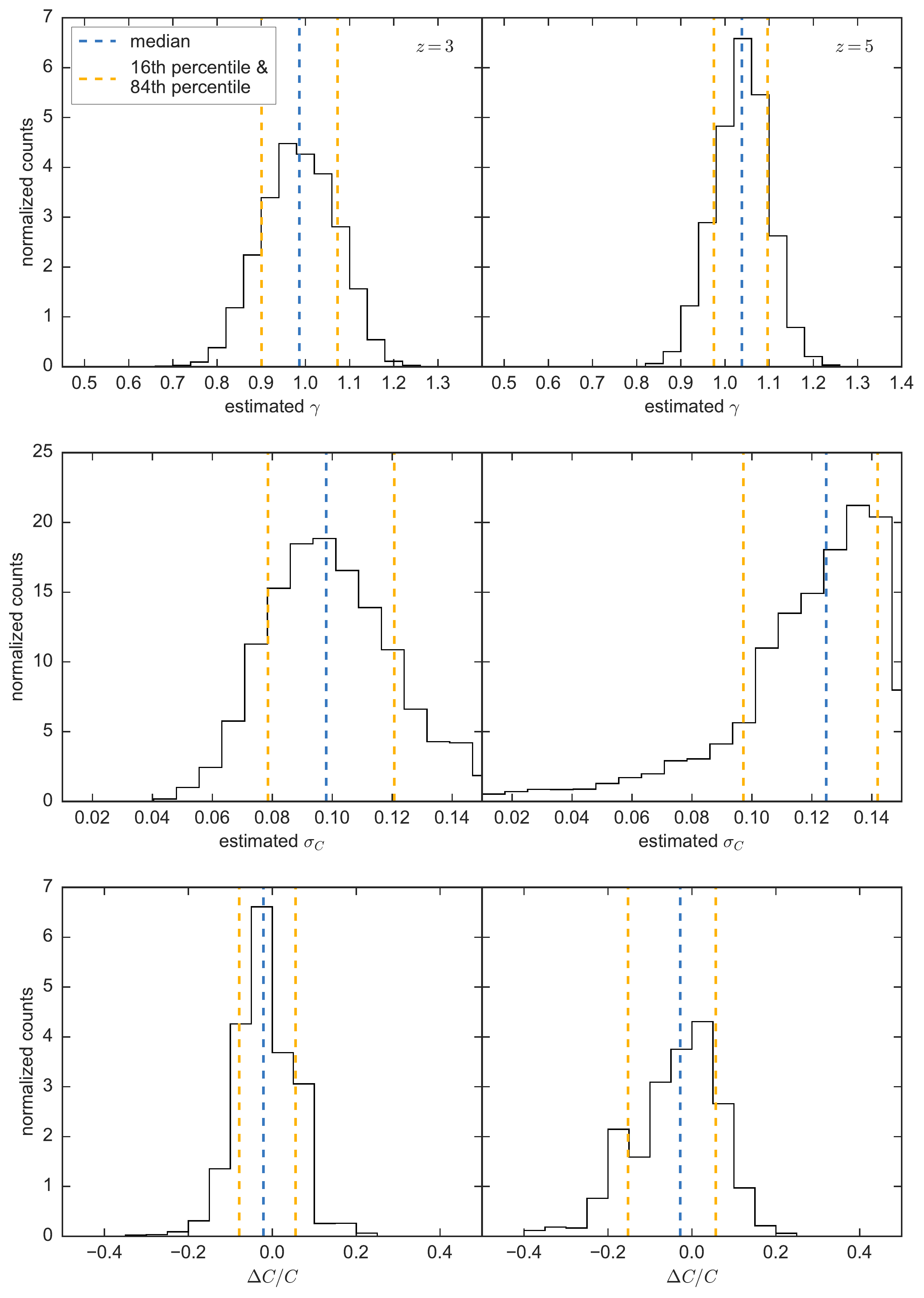}
\caption{Results from our MCMC algorithm for a realization of an ensemble of ten quasar spectra at redshift $z=3$ (left panels) and redshift $z=5$ (right panels). \textit{Upper panels:} posterior probability distribution of $\gamma$. The blue dashed lines show the median of the posterior
probability distributions and the yellow dashed lines represent the
$16$th and $84$th percentiles.  \textit{Middle panels:} posterior probability distribution of the estimated continuum error $\sigma_C$. \textit{Lower panels:} Distribution of continuum residuals $\Delta C/C$ in the wavelength region of the \lya forest of all ten spectra of the ensemble. \label{fig:results_method}}
\end{figure*}

We show the results for a realization of an ensemble of $N_{\rm
  QSO}=10$ quasar spectra and a continuum model with $N_{\rm PCA}=2$
in Fig.~\ref{fig:results_method}. We run our MCMC algorithm with $100$
walkers for $10,000$ steps and chose half the steps as the burn-in time
until convergence of the MCMC chains is achieved. The second half of the chains is used 
as posterior probability distributions. The upper panels show the
posterior probability distribution for the parameter $\gamma$ at
redshift $z=3$ and $z=5$ in the left and right panel,
respectively. In both cases the fiducial input parameter of $\gamma=1.0$ with which the
ensemble of mock quasar spectra was generated, is recovered. The width (i.e. $68\%$-region) of the posterior probability distribution, which gives the precision with
which we can recover $\gamma$ with this ensemble of ten quasar spectra, 
is $\pm8.6\%$ at redshift $z=3$ and $\pm6.1\%$ at $z=5$. Note that we quote the average between the ($50$th-$16$th)-percentile and the ($84$th-$50$th)-percentile here, since the distributions are fairly symmetric.

The middle panels show the posterior probability distribution of the continuum
error $\sigma_C$ in the estimation of the continuum model of all
quasar spectra in the ensemble. 
At redshift $z=3$ the ensemble of ten
quasar spectra that we chose in this run constrains the continuum
error to be $\sigma_C\approx 9.8\pm2.1\%$. 
At redshift $z=5$
the continuum error $\sigma_C$ is only mildly constrained. This is
not a surprising result given that the flux PDF at this high redshift is fairly insensitive to the
level of
uncertainty in the continuum estimation (see Fig.~\ref{fig:PDF_cont_error}). In this case the median value of $\sigma_C\approx12.5\pm2.2\%$
should be considered with caution since it will also be dependent on prior assumptions for $\sigma_C$. 

The lower panels of Fig.~\ref{fig:results_method} show the actual
continuum residuals $\Delta C/C$ for
all quasars in the ensemble calculated with eqn.~(\ref{eq:deltac}), i.e. it shows the relative continuum error between the true quasar continuum and the continuum model estimated via PDF regulation. The obtained median values of $\Delta C/C\approx
-2.1\%$ at $z=3$ and $\Delta C/C\approx-2.8\%$ at $z=5$ 
indicate that there is a very small systematic overestimation of the continuum level. The width of this residual distribution, the actual PDF regulated continuum error $\sigma_{\Delta C/C}$, i.e. the average of the ($84$th-$50$th) percentile and ($50$th-$16$th) percentile, shows that the actual continuum error for this ensemble of ten quasar
spectra is $\sigma_{\Delta C/C}\approx 6.7\%$ at $z=3$ and $\sigma_{\Delta C/C}\approx 10.5\%$ 
at $z=5$. These values are slightly lower than the estimated continuum error $\sigma_C$, i.e. $\sigma_C$ is slightly overestimated, but still within the $1\sigma$-region of the posterior probability distribution at $z=5$ and slightly outside of it ($1.46\sigma$) at $z=3$.

\begin{figure*}[t]
\centering
\includegraphics[width=.9\textwidth]{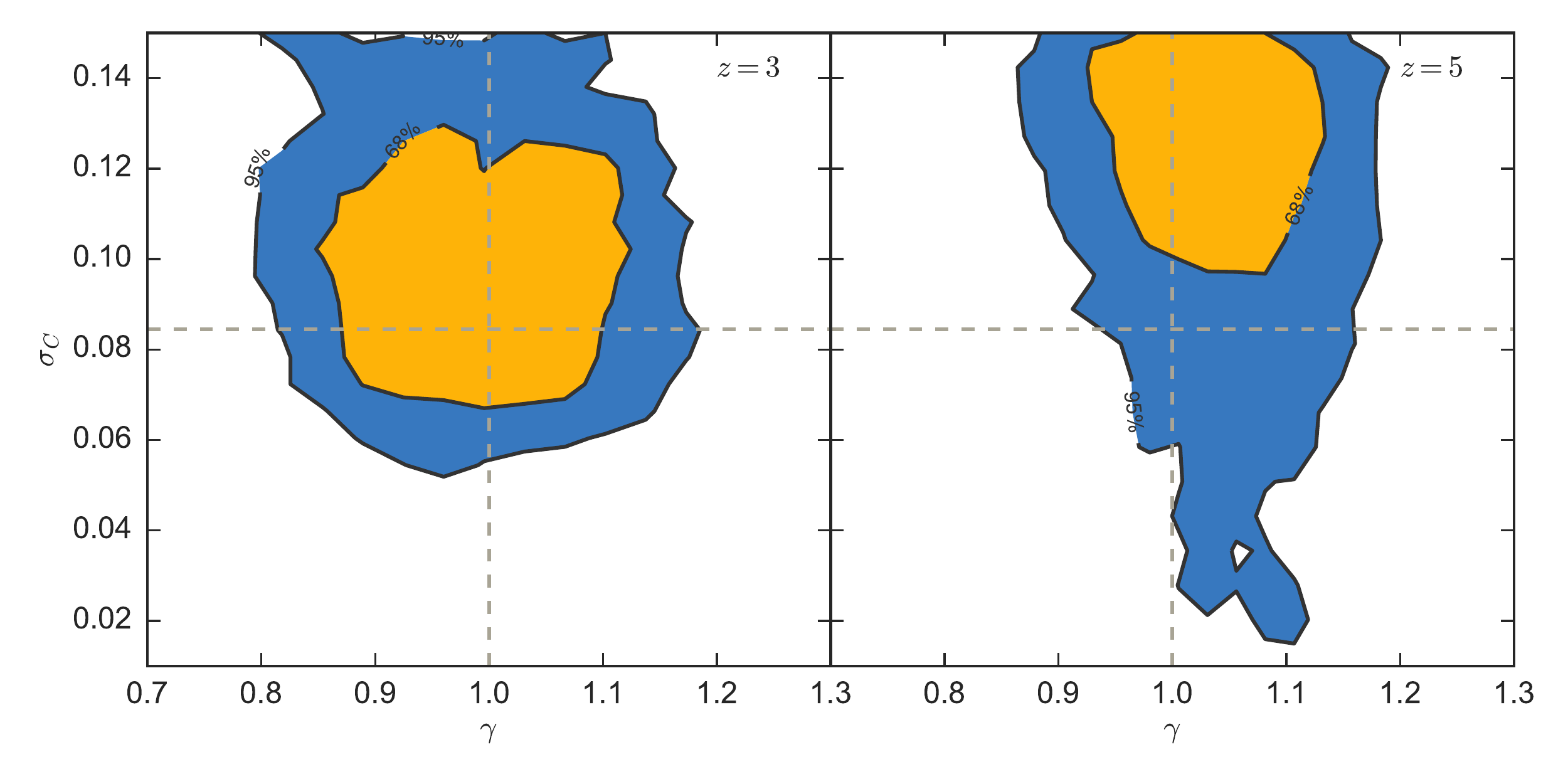}
\caption{Contour plot showing possible degeneracies of the two parameters $\gamma$ and $\sigma_C$ for an ensemble of ten quasar spectra at $z=3$ (left panel) and $z=5$ (right panel). The gray dashed lines indicate the fiducial input value of $\gamma=1.0$ and the continuum error $\sigma_{\rm PCA=2}\approx 8.4\%$ that is intrinsic to the continuum model with two PCS. \label{fig:contour}}
\end{figure*}

By means of the posterior probability distributions we can explore the
degeneracy between the two parameters that govern the shape of the
PDF. This is demonstrated in Fig.~\ref{fig:contour} at redshift $z=3$
and $z=5$ in the left and right panel, respectively. The slope parameter
$\gamma$ is shown on the x-axis, versus the 
continuum error $\sigma_C$ on the y-axis. The gray dashed lines
indicate the fiducial parameter of $\gamma=1.0$ and the intrinsic
error of the continuum model with two PCS, i.e. $\sigma_C=\sigma_{\rm PCA=2}\approx8.4\%$ (see Fig.~\ref{fig:residuals}).

At redshift $z=3$ we can obtain constraints for both parameters with our algorithm, since there is no degeneracy between the two parameters of interest. This result can be seen directly in Fig.~\ref{fig:pdf_different_gammas} and Fig.~\ref{fig:PDF_cont_error}, since $\gamma$ and $\sigma_C$ alter the shape of the PDF in different manners. 
At higher redshift at $z=5$ the constraints on $\sigma_C$ are very weak, because the PDF of the transmitted flux in the \lya forest is less sensitive to this parameter as we have seen in Fig.~\ref{fig:PDF_cont_error}. However, at both redshifts we can constrain the slope parameter $\gamma$ of the temperature-density relation of the IGM, which was our main goal.

\subsection{Robustness of the PDF Regulated Continuum Fitting}\label{sec:res_all}

In this subsection we investigate the robustness and reliability of
our method for ensembles of ten high-resolution quasar spectra. Hence
we would like to determine whether our method results in possible
biases in the model parameters, whether the estimated continuum error
$\sigma_C$ is a good proxy for the actual PDF regulated continuum
error $\sigma_{\Delta C/C}$, and how well the continuum error compares
to the intrinsic accuracy of the chosen PCA basis.  Therefore we run
our algorithm a hundred times, each time with a different realization
of an ensemble of ten quasar spectra. Each of the $100$ runs entails
an MCMC, which results in posterior probability distributions for
$\gamma$ and $\sigma_C$ as well as for the coefficients of the PCS for
each quasar. Taking the mean of each posterior probability
distribution of the PCS gives us a mean continuum model for each
quasar spectrum, which enables us to calculate the continuum residual
distribution $\Delta C/C$ and its width, the actual PDF regulated
continuum error $\sigma_{\Delta C/C}$.

\begin{figure*}
\centering
\includegraphics[width=.9\textwidth]{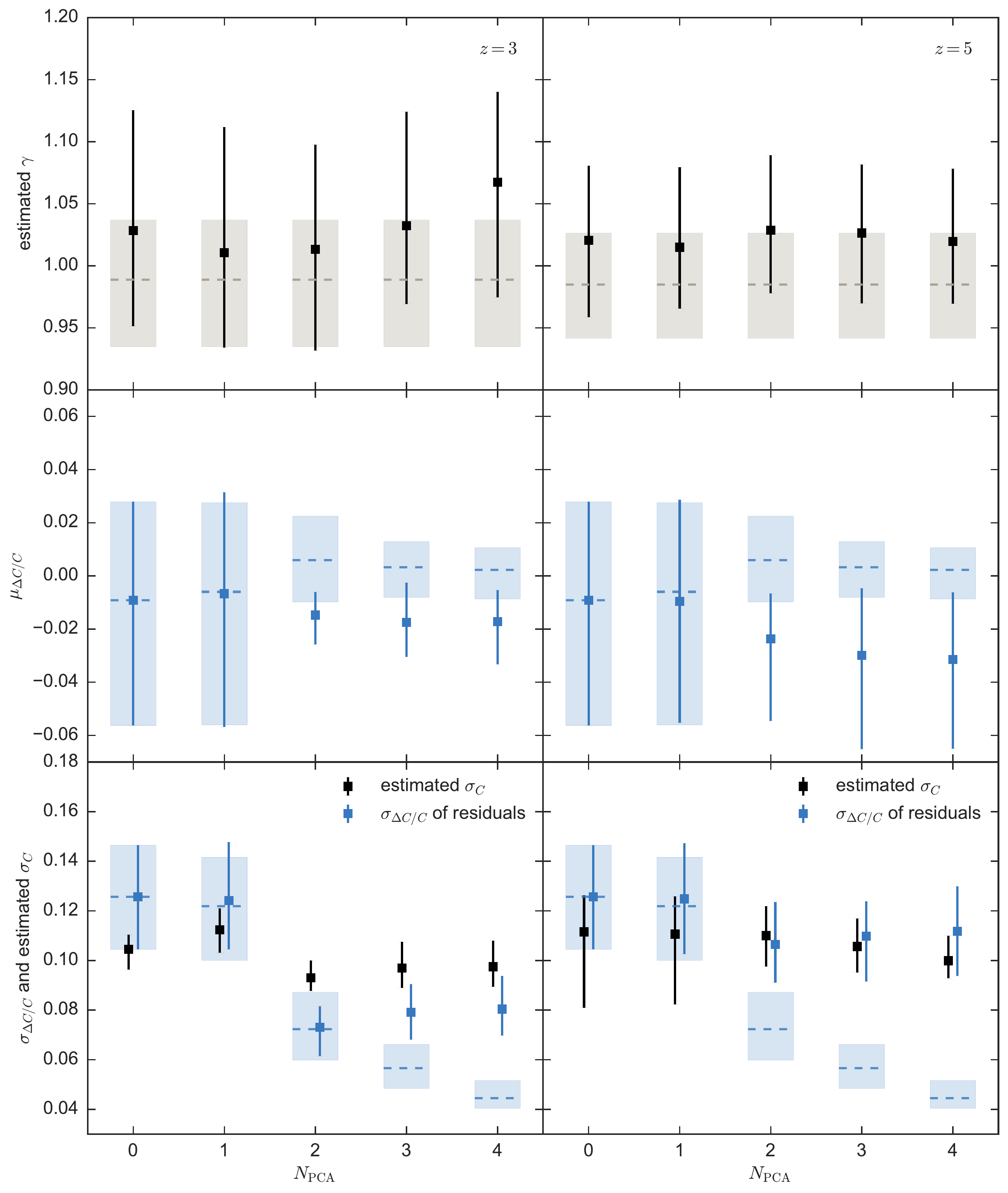}
\caption{Estimated values for $\gamma$ (upper panels), the continuum residuals (middle panels) and the continuum error (lower panels) dependent on the number of PCS that were taken into account to model each quasar continuum at redshift $z=3$ (left panels) and redshift $z=5$ (right panels). Shown are the median and $16$th and $84$th percentiles. \textit{Upper panels:} distribution of estimates for $\gamma$ with gray shaded areas showing the ideal distribution of estimates for $\gamma$ when noise-free quasar spectra and perfect knowledge of the quasar continua are assumed. Note that the fiducial value is $\gamma=1.0$, but since we only take 100 realizations the distribution fluctuates around this fiducial value. In this case $\gamma$ is the only free parameter to estimate from the ensemble. 
  For each data point we run our algorithm a hundred times with a different realisation of the ensemble of ten quasar spectra and plot the distribution of the median values of the posterior probability distributions. \textit{Middle panels:} distribution of medians of the continuum residual distribution in the \lya forest region. The blue shaded areas show the continuum residual distribution intrinsic to the continuum models. \textit{Lower panels:} showing the continuum errors; the blue data points indicate the real measured widths $\sigma_{\Delta C/C}$ of the continuum residuals whereas the black data points show the distribution of median estimates of the inferred continuum error $\sigma_C$. The blue shaded areas show the intrinsic error of the continuum model due to the limited number of principal components. }\label{fig:results_inf}
\end{figure*}

The results of this investigation are presented in Fig.~\ref{fig:results_inf} for redshift
$z=3$ and $z=5$ in the left and right panels, respectively. 
The upper panels show the 
estimated values for the slope parameter
$\gamma$ of the temperature-density relation
as a function of $N_{\rm PCA}$
that were used to model the continuum of each
quasar. From each of the hundred runs we obtain a posterior probability
distribution for $\gamma$ and adopt its median as the best estimate (see upper panels of Fig.~\ref{fig:results_method}). We
then examine this distribution of estimates of $\gamma$ from the hundred realizations of the quasar ensemble. The data
points in the upper panels of Fig.~\ref{fig:results_inf} indicate
the median of this distribution, i.e. the $50$th percentile, and the
errorbars show the $16$th and $84$th percentile. In order to compare to the ideal but unrealistic situation when assuming ensembles of ten noise-free spectra with no continuum error
we also show the gray areas that show the median (gray dashed lines) and the $16$th to $84$th percentile
region of estimates of $\gamma$. In these cases $\gamma$ is the only free parameter to
estimate with MCMC.

These upper panels of Fig.~\ref{fig:results_inf} show that we can
recover the thermal parameter $\gamma$ at both redshifts with an
ensemble of ten quasar spectra remarkably well. The medians of the
posterior probability distributions for $\gamma$ are mostly within the
$16$th to $84$th percentile region of the estimates for ensembles of
perfect noise-free spectra, and the fiducial input value of $\gamma=1.0$ lies within the $68\%$-region of all estimates.  
For $N_{\rm PCA}=2$ which we will adopt as our fiducial model, the median estimate for $\gamma$ and its $1\sigma$ fluctuations at $z=3$ are $\gamma\approx 1.013$ with $\sigma_{\gamma} = 8.3\%$ and $\gamma\approx 1.029$ with $\sigma_{\gamma} = 5.6\%$ at $z=5$. We expect 
$1\sigma$-fluctuations around the fiducial value of $\gamma=1.0$ of $\sigma_{\rm med \gamma}\simeq 8.3\% /\sqrt{100}\approx0.83\%$ at $z=3$ ($\sigma_{\rm med \gamma}\simeq 5.6\% /\sqrt{100}\approx0.56\%$ at $z=5$) 
 due to the fact that we only investigate $100$ different realizations of our ensemble. Hence the estimated values of $\gamma$ lie marginally outside the expected region. 
This may be a random fluctuation, however, since the other estimates of $\gamma$ for different values of $N_{\rm PCA}$ all fluctuate in the same direction it is suggestive that the values are slightly biased. Nevertheless the bias in the estimation of $\gamma$ is still $6-7$ times smaller at $z=3$ ($2-3$ times smaller at $z=5$) than its $1\sigma$ errorbars and thus we can tolerate this possible small bias.
The marginalization over noise and continuum error leads to a modest reduction of precision in the measurement of $\gamma$, since the errorbars increase by a factor of $\sim 2$ compared to the ideal noise-free case (gray areas). 

The middle panels of Fig.~\ref{fig:results_inf} illustrate how
centered the continuum residuals obtained by PDF regulation are about
zero.  For each MCMC inference we compute the continuum residuals
$\Delta C/C$ in the region of the \lya forest (i.e. rest-frame
wavelengths $1040${\AA}$\leq\lambda\leq 1190${\AA}) with
eqn.~(\ref{eq:deltac}), and compute the median $\mu_{\Delta C/C}$ of
the distribution.  This results in a distribution of 100 median values
$\mu_{\Delta C/C}$, and the median of this distribution of medians (points),
as well as its $16$th and $84$th percentiles (errorbars), are plotted
in the middle panels. The blue dashed lines and
shaded regions indicate the corresponding $\mu_{\Delta C/C}$ resulting
from errors intrinsic to the PCA continuum model. Specifically, we
MCMC fit only the continua of the the same 100 realizations of $N_{\rm
  QSO}=10$ quasars with the PCA model, and performed exactly the same
procedure on the resulting $\Delta C/C$ distributions, and 100
$\mu_{\Delta C/C}$ median values.

A completely unbiased continuum estimation would give a continuum
residual distribution with a median of $\mu_{\Delta C/C}\approx0$. We
see that both at $z=3$ and at $z=5$, there is a small negative bias of a few percent in the estimates, i.e. we slightly overestimate the
quasar continuum level. However, this bias is below $2\%$
at $z=3$ and below $3\%$ at $z=5$ for $N_{\rm PCA}=4$ and even
smaller for $N_{\rm PCA}<4$, and thus insignificant relative to the
typical continuum residuals $\Delta C\slash C$ of $\sim 7-10\%$
(lower panel Fig.~\ref{fig:results_method}).
For $N_{\rm
  PCA}=2$, which represents our fiducial model, our algorithm
overestimates the continuum by $\approx1.5\%$ at $z=3$ and
$\approx2.4\%$ at $z=5$. Our measured $\mu_{\Delta C/C}$ are
marginally consistent, i.e. the $68\%$-regions overlap, with the
continuum residuals intrinsic to the continuum model, i.e. the shaded
blue regions. For $N_{\rm PCA}\geq 2$ our PDF regulated continuum fits
are just slightly more biased than the one obtained by fitting the PCA model
to the quasar spectra in the absence of Ly$\alpha$ forest.
However, the biases we obtain are much smaller
than those estimated for hand-fitted quasar continua \citep[see
  e.g.][]{FaucherGiguere2008}.

The lower panels of Fig.~\ref{fig:results_inf} illustrate the behavior
of the $1\sigma$ error of our continuum estimates. Recall that
we characterize
continuum errors in two different
ways: once 
we fit for the continuum error $\sigma_C$ and obtain an estimate via MCMC; in the other case we calculate the continuum residuals $\Delta C/C$ using our knowledge of the
the true continua of the mock spectra, and take the width of this distribution $\sigma_{\Delta C/C}$ as the estimate for the continuum error. 
The black points with error bars represent the median and the $16$th
and $84$th percentiles of the inferred continuum error $\sigma_C$ from
our MCMC PDF continuum regulation. Specifically, for each of the 100
realizations of ten quasar spectra, we performed MCMC PDF regulation, obtained a posterior probability distribution for $\sigma_C$ and
adopted its median value as the best estimate. This results in 100
median $\sigma_C$ values, where the median and the $16$th and
$84$th percentiles (errorbars) are shown in black. 
The blue data
points correspond to the actual PDF regulated continuum error, i.e.
the measured dispersion $\sigma_{\Delta C/C}$ of the continuum residuals
$\Delta C/C$. For each run we compute the continuum residuals $\Delta C/C$ for all pixels for the whole ensemble of quasars and take the average of the $16$th and $84$th percentile of this distribution as the estimate for $\sigma_{\Delta C/C}$. We obtain 100 of these measurements for each realization of the quasar ensemble and plot the median value and the $16$th and $84$th percentiles. 
The blue shaded areas show the same distributions of median values for
the ideal case with the noise-free spectra and no \lya forest,
i.e. the error intrinsic to the continuum model for a given $N_{\rm
  PCA}$
\footnote{This corresponds to the lower panel of
Fig.~\ref{fig:residuals} with the difference being that for the
previous plot we calculate the width of the continuum residuals for
all $50$ HST quasars, whereas in Fig.~\ref{fig:results_inf} we show
the distribution of median values of the widths of the continuum
residuals from each ensemble of ten quasars. Thus the blue dashed
lines in Fig.~\ref{fig:results_inf} and the data points in the lower
panel of Fig.~\ref{fig:residuals} do not agree exactly but differ
slightly due to random noise fluctuations. }.

Comparing the actual error of the PDF regulated continuum residuals
(blue points) with the intrinsic precision of the PCA basis (blue
shaded region) in the lower left panel at $z=3$, one sees
that our algorithm recovers the continuum consistent with the best accuracy
achievable with the PCA basis, at least for $N_{\rm PCA} \leq 2$. 
For higher values of $N_{\rm PCA} > 2$ the blue points lie slightly
above the intrinsic accuracy of the PCA model shown by the shaded
regions, and thus our PDF regulated continuum fits perform slightly
worse than the underlying PCA model. One might expect the PDF
regulated continuum error (blue points) to continue to decrease as
$N_{\rm PCA}$ is increased, following the intrinsic error of the PCA
basis (shaded region), but instead the PDF regulated continuum
errors saturate around a value of $\sigma_{\Delta C\slash C}=0.07$.
We believe that this is due to the floor that we introduced into our covariance matrix (see \S~\ref{sec:covariance})
in order to prevent it from becoming singular. This floor sets a
threshold below which we can no longer measure continuum errors. Once
the chosen PCA basis gets accurate enough, we can no longer recover
the small continuum errors, since the floor reduces the sensitivity
particularly in the high flux bins of the PDF, which are most sensitive to the
continuum error (see Fig.~\ref{fig:PDF_cont_error}).

Our results at $z=5$ in the lower right panel of
Fig.~\ref{fig:results_inf} tell a similar story.  The PDF regulated
continuum error (blue points) do not follow the intrinsic error of the
PCA basis (shaded region) for $N_{\rm PCA} \geq 2$. This most likely results
from the reduced sensitivity of the PDF at $z =5$ to the continuum (see Fig.~\ref{fig:PDF_cont_error}), and possibly also due to the floor we imposed in the covariance
matrix.

A comparison between the black data points showing the inferred
continuum error $\sigma_C$ and the blue data points indicating the PDF
regulated continuum error $\sigma_{\Delta C/C}$ at $z=3$ reveals that our
algorithm gives a fairly good estimate for the actual continuum
error. Nevertheless the inferred values are slightly biased. We
underestimate the continuum error for $N_{\rm PCA}=0$ and $N_{\rm
  PCA}=1$ and slightly overestimate it for $N_{\rm
  PCA}\geq2$. However, the inferred values of $\sigma_C$
usually lie within the $16$th and $84$th percentile region of the
real measured value $\sigma_{\Delta C/C}$ .
The errorbars on the $\sigma_C$ estimates are very small indicating
that the MCMC always converges to the same range of values. We believe
that this is also an artifact resulting from the floor imposed on our
covariance matrix.

At redshift $z=5$ the algorithm seems to recover the actual
continuum error better, i.e. the inferred
continuum error $\sigma_C$ (black data points) and real measured PDF regulated continuum error $\sigma_{\Delta C/C}$ (blue data points)
agree fairly well, but since the locations of the black inferred
values are not well constrained because the posterior probability
distributions for $\sigma_C$ is fairly flat (see middle right panel of
Fig.~\ref{fig:results_method}), this agreement is partly coincidental and
dependent on our choice of the prior probability distribution on $\sigma_C$ in the MCMC.

In summary, the lower panel of Fig.~\ref{fig:results_inf} illustrates
that
our PDF regulated continuum fitting method is producing continuum errors
at $z=3$ that roughly track the intrinsic precision of the PCA basis,
since both the actual PDF regulated continuum error $\sigma_{\Delta C/C}$ (blue points)
as well as the inferred continuum error $\sigma_C$ (black points)
lie within the blue shaded region for $N_{\rm PCA} \leq 2$. At redshift $z=5$ the inferred continuum uncertainties are only weakly constrained but nevertheless reliable.

How many PCS should be used?  At $z = 3$, the
lower left panel shows a drop in the continuum error in both the inferred continuum error $\sigma_C$ and the real measured continuum error $\sigma_{\Delta C/C}$ 
and
thus an increase of the precision in the continuum model when two or
more principal components are taken into account. At redshift $z=5$
the drop in the continuum error is small, which is due to the fact
that the PDF at this redshift barely constrains the quasar continuum
level as we have seen before in Fig.~\ref{fig:PDF_cont_error}. At both
redshifts the gains in precision for more $N_{\rm PCA}$,
i.e. $N_{\rm PCA}=3$ or $N_{\rm PCA}=4$, is insignificant. 
Since we are only interested
in the continuum residuals in the wavelength region of the \lya
forest, only the PCS that alter the shape of the quasar continuum in
this region are influential for our method. The first principal
component spectrum for example mainly addresses the shape of the \lya
emission line, and hence does not improve the continuum estimation in
the \lya forest region. This is why the continuum error is not reduced
before taking at least two PCS, i.e. $N_{\rm PCA}\geq 2$, in the
continuum model.

Thus we advocate using the continuum model with two PCS for this
analysis, since we want to minimize the number of free parameters and
thus the dimensionality of our parameter space which in turn limits
the computational time. This implies that we have two nuisance
parameters for the continuum estimation per quasar plus the global
parameters that influence the shape of the PDF --- $\sigma_C$ and any
parameters related to the physics of the IGM, which for our simplified
IGM model is only $\gamma$.  For an ensemble of ten quasar spectra this
results in a $2\times N_{\rm QSO}+2=22$ dimensional parameter
space. However, in principle our method can accommodate larger
ensembles of quasar spectra and a more complicated model of the IGM.

\section{Discussion and Caveats}\label{sec:discussion}

In this section we will discuss several caveats of our algorithm and possibilities for further development that will be subject to future work. 

\subsection{Enlarging the Parameter Space}\label{sec:caveats_parameterspace}
Although this study only considered a highly simplified single
parameter model of the thermal state of the IGM, in practice our
method can easily accommodate a more complicated multi-parameter PDF
model, the only limitations being computational time and effort. When
applying this method to real quasar spectra one would use
hydrodynamical simulations to generate the IGM model and also consider
the temperature at mean density $T_0$, the Jeans pressure smoothing
scale $\lambda_{\rm P}$ \citep{rorai2013, Kulkarni2015}, and other possible
nuisance parameters \citep{LeeHennawi2015}. Adding more thermal
parameters or other nuisance parameters related to the IGM (e.g. Lyman
Limit Systems) extends the computational effort only marginally. More
parameters in the continuum model on the other hand, add a more
significant amount of time to the computation because they need to be
estimated for each quasar in the ensemble and hence the number of
continuum parameters multiplies with $N_{\rm QSO}$. In this work we
compressed the continuum error of each quasar spectra into only two
nuisance parameters. We tested our algorithm with an ensemble of ten
quasar spectra, but a larger ensemble with even double the number of
quasar spectra is still easily feasible.  A larger ensemble of quasar
spectra would also further improve the precision on our estimates of
the thermal parameters.

\subsection{Fixing the Mean Flux}

The mean flux of the \lya forest has been held fixed in our analysis. Therefore, we computed a hundred different realizations of the \lya forest for each value of $\gamma$ that we considered in our models and computed a scaling factor for the optical depth $\tau$ averaged over the hundred \lya realizations, in order to match the mean flux in our mock spectra to measurements of the mean flux (see \S~\ref{sec:mock_lya_forest}). While creating each mock quasar spectrum we re-scale $\tau$ with the scaling factor. In this way, the average mean flux of the \lya forest is kept fixed, but individual mock spectra can have higher or lower mean flux values, which is expected to be the case also for real quasar spectra due to cosmic variance.

The mean flux could be considered as an additional nuisance parameter in the PDF models. This would not add  significantly to the computation time (see \S~\ref{sec:caveats_parameterspace}), but might decrease the precision of the thermal parameter estimation slightly. However, given the precision of the most recent mean flux measurements ($\Delta \langle f\rangle_{z\sim 3}\approx 1\%$ and $\Delta \langle f\rangle_{z\sim 5}\approx 7\%$ \citet{becker2013}), we opted to keep it fixed.

\subsection{Redshift Evolution in the \lya Forest}

In our simplified model for \lya forest absorption we do not take the redshift evolution within the \lya forest along the line of sight into account, i.e. we assume a fixed value of $\gamma$. However, assuming a redshift evolution within the \lya forest and thus within the temperature-density relation of the IGM, would result into a redshift dependent slope parameter, i.e. $\gamma(z)$. Our parameter space of the PDF models can be easily augmented with a redshift dependent $\gamma$ and we would only need to account for one more free parameter in the MCMC runs. 

\citet{Rorai2016} discuss temperature-density relations with a break in the slope, i.e. a different slope for over- or underdensities. Even such non-linear dependencies can be easily accommodated in our algorithm by simply constructing a different set of PDF models that contain two different linear slope parameter $\gamma$. These models can be parameterized by two parameters $\gamma_{\rm o}$ and $\gamma_{\rm u}$ for over- and underdense regions, respectively, which simply adds another free parameter to our algorithm. 

In principle the PDF models can be made arbitrarily complicated by adding more and more nuisance parameters that will be marginalized out in the end.

\subsection{The Choice of the PCA Basis}
A more suitable set of PCS could further increase the precision of the
continuum model and reduce the continuum error without increasing the
number of free parameters in the continuum model and thus the
computational time. The PCS that we use in this work from
\citet{Suzuki2006} cover the the whole spectral range from
$1020${\AA}-$1600${\AA} of quasar spectra. However, for the studies of
the IGM, one is interested in a small range of wavelengths, i.e. the
spectral range spanning the \lya forest.  Thus some of the PCS we use
account for the different shapes in the Ly$\alpha$ and other emission
lines for example, but do not help improve the continuum model in the
\lya forest region. One could thus come up with a different PCA basis
that would be more suited for our purposes. Other methods, such as an
Independent Component Analysis \citep[e.g.][]{Allen2011} or Heteroscedastic Matrix Factorization \citep[e.g.][]{Tsalmantza2012}, could also
be useful to construct a more suitable continuum model while keeping the dimensionality of the PCA low. 

Once applying this algorithm to real quasar spectra, the chosen set of PCS might not be flexible enough to account for enough of the variance in the spectra. However, for high-z spectra, the PCA basis of \citet{Paris2011} could be applied for instance, or a new set of PCA could be created from a similar procedure by fitting the $> 10,000$ spectra available with high $\rm S/N$ ratio in the BOSS survey. If such an augmented and improved PCA were constructed and fit to real data, if the red side of each spectrum is modeled well by the PCA, there is no reason to believe that the blue side is not expected to match.

\subsection{Construction of Covariance Matrices}

We used an approximation when constructing the covariance matrix,
whereby we fixed the covariance matrix to have the intrinsic continuum
error of a single continuum model. This fact required us to add a
floor to the covariance matrix according to the matrix shrinkage
approach, in order to prevent the matrix from becoming singular, since
the very high and very low flux bins were rarely populated for the fiducial
model used to construct the covariance.
This approximation set a floor on the accuracy of our PDF regulated 
continuum, and limited our ability to obtain more precise continuum fits
by increasing the number of PCS. 
Thus in the future
it would be worthwhile to construct covariance matrices as a function
of the model parameters $\gamma$, $\sigma_C$ and possible others. This
would remove the bias in the continuum residuals and increase the
precision of the continuum estimation. However, constructing
covariance matrices with varying $\sigma_C$ is tricky, since we
introduced the continuum error into our PDF models by construction
without any covariance (eqn.~(\ref{eq:flux})). One possible approach
to solve this problem could be to interpolate between the covariance
matrices with continuum errors intrinsic to the chosen continuum model. 
This will be the subject of future work. 

\subsection{Flux Calibration of the Quasar Spectra}

Applying our new algorithm to real quasar spectra requires the spectra
to be flux calibrated. \citet{Suzuki2003} showed how one can flux
calibrate Echelle spectra using lower resolution spectra with accurate
spectrophotometry. We do not expect any problems arising due to errors
and uncertainties on this flux calibration, since we infer the
continuum error $\sigma_C$ with MCMC, and any flux calibration errors
would become a part of the continuum error that we can marginalize
out.

\section{Summary and Conclusion}\label{sec:summary}

We presented a new Bayesian algorithm making use of MCMC sampling that
allows us to simultaneously estimate the unknown continuum level of
each quasar in an ensemble of high-resolution spectra as well as their
common \lya forest flux PDF. This fully automated PDF regulated
continuum fitting method models the unknown quasar continuum with a
PCA basis with the coefficients of the principal components treated as
nuisance parameters. This method allows us to estimate parameters
governing the thermal state of the IGM, such as the slope of the
temperature-density relation, while marginalizing out
continuum uncertainties in a fully Bayesian way.

The primary results of this study are: 
\begin{itemize}
\item The intrinsic error in the model for the quasar continuum
  due to a finite number of PCS has a
  significant impact on the shape of the PDF of the transmitted flux
  in the \lya forest at $z=3$, whereas the \lya flux PDF at
  higher redshifts of $z=5$ is mostly unaffected by the continuum
  error. 
  
  \item We incorporate the continuum error
  and Gaussian white noise into the models for the PDF and into the
  covariance matrices and show that by treating these effects as nuisance parameters and marginalizing out the continuum error and noise our algorithm removes any biases in the estimation of the
  thermal parameters.

\item We show that our algorithm recovers $\gamma$, the only thermal parameter we consider in a simplified model of the IGM, without a significant bias and with high precision for an ensemble of ten high-resolution quasar spectra. We obtain a precision of $\pm 8.6\%$ at $z=3$ and $\pm 6.1\%$ at $z=5$ marginalized over all uncertainties in the continuum estimation. 

\item We explore the degeneracy between the two free parameters of our PDF models, the thermal parameter $\gamma$ and the model dependent parameter for the continuum error $\sigma_C$. As expected we do not see strong degeneracies since both parameters alter the shape of the PDF in slightly different ways. At $z=5$ the constraints on $\sigma_C$ are very weak as we expect from the insensitivity of the PDF to this parameter at high redshifts.

\item We advocate using $N_{\rm PCA}=2$ for modeling each quasar continuum since it minimizes the number of continuum parameters without significant loss in precision when determining thermal parameters or the quasar continua. 
\item The PDF regulated continuum achieves a precision of $\sigma_{\Delta C/C}\approx 6.7\%$ at $z=3$, which is in agreement with the intrinsic precision of the continuum model. At $z=5$ given the limited sensitivity of the \lya flux PDF, we cannot quite recover the intrinsic precision of the continuum model. The PDF regulated continuum at $z=5$ has an error of $\sigma_{\Delta C/C}\approx 10.5\%$.

\item At $z=3$ we show that our estimated continuum error $\sigma_C$ for an ensemble of ten quasar spectra tracks the underlying real measured continuum error fairly well for $N_{\rm PCA}\leq 2$. At $z=5$ the PDF of the \lya forest flux is only slightly dependent on the continuum estimation and thus we only obtain mild constraints for the continuum error $\sigma_C$ that nevertheless give a rough estimate of the actual underlying continuum error.  
 
\item The distribution of continuum residuals shows that the quasar continua are generally slightly overestimated by $\mu_{\Delta C/C}\approx 1.5\%$ at $z=3$ and $\mu_{\Delta C/C}\approx 2.4\%$ at $z=5$ for a continuum model with $N_{\rm PCA}=2$. 
\end{itemize}

Our algorithm improves upon previous work in two important ways.
First, the method is totally automated and thus avoids
tedious manual fitting of the quasar continua. Fitting quasar continua
by hand could result in continuum errors that are correlated with flux
and dependent on the underlying IGM model \citep{Lee2012}, which results
in systematic errors that are very difficult to model.  Secondly, we
model the errors due to imprecise continuum fits and treat them
as a nuisance parameter. This results in unbiased estimates of the
thermal parameters that are our primary interest, with all
continuum uncertainties marginalized out in a Bayesian way. 
Given the large amount of high-resolution and high $\rm S/N$ quasar data now publicly available \citep{KODIAQ, OMeara2015}, the time is ripe to apply this new methodology to real data.

\section*{Acknowledgment}
We thank Alberto Rorai, David W. Hogg and the members of the ENIGMA group\footnote{http://www.mpia-hd.mpg.de/ENIGMA/} at the
Max Planck Institute for Astronomy (MPIA) for helpful discussions. 

We would also like to thank the referee for his helpful and thorough comments that significantly improved this paper.

\bibliography{literatur}


\appendix

\section{The Influence of the Spectral Coverage on the Red Side of the \lya Emission Peak}

So far we only considered the part of the quasar spectrum on the blue side of the \lya
emission line, i.e. at shorter wavelengths, where the \lya forest is located. However, observed quasar spectra
usually cover a much larger wavelength range and extend also to the red side
of the \lya emission line, i.e. larger wavelengths, where only a few metal lines but no
hydrogen absorption interrupt the continua. Therefore this
wavelength range might contain information that could improve our
estimates of the continuum level in the Ly$\alpha$ forest region and
hence also the constraints on thermal parameters. 

There are two questions we would like to
address here: how can we use the information on the red
side of the quasar spectrum for the fit of the PDF of the transmitted
\lya flux and the thermal parameter estimation? And how much
information about the continuum shape in the \lya forest region is contained at
wavelengths redwards of the \lya line? Therefore we will
first introduce a second likelihood function that determines the
quasar continuum on the red side of each spectrum. We will then
combine this new likelihood function with the previous one from
eqn.~(\ref{eq:L_PDF}) to form a joint likelihood function and evaluate
at the end of this section the impact of adding this information to
the estimation of the thermal parameter $\gamma$.

\subsection{The Likelihood Function for the Continuum on the Red Wavelength Side of the Quasar Spectra $\mathcal{L_{\rm cont}}$}\label{sec:lcont}

In \S~\ref{sec:pca_cont_model} we have introduced a likelihood function $\mathcal{L}=\exp(-\chi^2/2)$ (see eqn.~(\ref{eq:chi2})) to find estimates for the continuum model to the data that we used to evaluate the intrinsic continuum error of the model. We will now use this likelihood function for the continuum model of each quasar on longer wavelengths than the \lya emission line, i.e. on the red side of the spectrum. 
The likelihood function now reads:
\begin{align}
\mathcal{L_{\text{cont}}} &= \mathcal{L}(\alpha_{ij}|C_{\text{data}})\nonumber\\
&=\prod_i^{N_{\text{QSO}}}\prod_{\lambda}\frac{1}{\sqrt{2\pi\sigma_i^2}}\exp\left(-\frac{(C_{\text{model,} i\lambda}(\alpha_{ij})-C_{\text{data,} i\lambda})^2}{2\sigma_{i, \rm noise}^2}\right). \label{eq:L_cont}
\end{align}
The free parameters of the continuum model $\alpha_{ij}$ are the coefficients of the PCS $j$ of each quasar $i$ as previously seen in eqn.~(\ref{eq:pca}). Thus the likelihood function evaluates how well the continuum model given by eqn.~(\ref{eq:pca}) describes the data on the red side of each quasar spectra. 

It is important to note that only the wavelength range \textit{redwards} of the \lya emission line will be taken into account to evaluate this likelihood function, i.e. $1216${\AA}-$1600${\AA},
since there are no absorption lines present (we neglect possible contamination due to metals that are not incorporated into our mock data set --- in real data these metals would be straightforward to mask.), whereas the wavelength region \textit{bluewards} of \lya will be considered by the previous likelihood function from eqn.~(\ref{eq:L_PDF}) as before. 

Thus $\mathcal{L_{\rm cont}}$ takes every pixel at wavelengths on the red side of the quasar spectrum into account, i.e. $\sim40,000$ data points. This likelihood function is clearly very different in character compared to the previous likelihood function $\mathcal{L_{\text{PDF}}}$ (eqn.~(\ref{eq:L_PDF})), where we bin the data and evaluate the difference between the data and the model for every bin in the PDF and thus the likelihood function $\mathcal{L_{\text{PDF}}}$ only covers $\sim25$ data points. In the next subsection we will combine the two likelihood functions $\mathcal{L_{\text{PDF}}}$ and $\mathcal{L_{\text{cont}}}$ and will see, that we have to slightly modify this new likelihood function $\mathcal{L_{\text{cont}}}$ due to its differences in character in order to make the two likelihood functions compatible. 

\subsection{Combining the Two Likelihood Functions}

We would like to join the two likelihood functions \lcont and \lpdf in order to make use of the information on the continuum level provided by the wavelength coverage on the red side of the \lya emission peak, but nevertheless regulate the continuum level via the \lya forest flux PDF on the blue side of each quasar spectrum. However, joining these two likelihood functions by multiplication is not as easy as one would naively expect. We have already mentioned that the character of the two likelihood functions is quite different: in one case we evaluate the likelihood 
for each pixel in the wavelength range between $1216${\AA}-$1600${\AA}, i.e. $\sim 40,000$ data points,
whereas in the other case we only evaluate the likelihood after binning the pixels, which result in more than three orders of magnitudes less data points. 
We know that the mean of a $\chi^2$ distribution is roughly equal to the number of degrees of freedom, which corresponds to the number of our data points. Since both likelihood functions are of the shape $\mathcal{L}=\exp(-\chi^2/2)$, this implies that the new likelihood function is much steeper than the other. Hence in a simple multiplication of the two likelihood functions, the steeper function will always dominate the joint likelihood function. 

We will therefore modify the new likelihood function $\mathcal{L_{\text{cont}}}$ in order to make the two likelihood functions more compatible. We will augment the function with an additional uncertainty $\sigma_{\rm red}$, which reflects the inability of the red side of each spectrum to predict the continuum on the blue side. Note that this parameter is not one that we estimate with MCMC but rather one that we infer by conducting various tests. 

The likelihood function \lcont then reads:
\begin{align}
\mathcal{L_{\text{cont}}}&= \mathcal{L}(C_{\text{model}}(\alpha_{ij})|C_{\text{data}}, \sigma_{\rm red})\nonumber\\
&= \prod_i^{N_{\text{QSO}}}\prod_{\lambda}\frac{1}{\sqrt{2\pi\sigma_i^2}}\exp\left(-\frac{(C_{\text{model,} i\lambda}(\alpha_{ij})-C_{\text{data,} i\lambda})^2}{2(\sigma_{i, \rm noise}^2 \color{red}{+ \sigma_{\rm red}^2}\color{black})}\right). \label{eq:L_cont_new}
\end{align}
Now the two likelihood functions can be joined together by simply multiplying them:
\begin{align}
\mathcal{L}(\gamma, \sigma_C, \alpha_{ij}) = \mathcal{L_{\text{cont}}}(\alpha_{ij}|\sigma_{\rm red})\times\mathcal{L_{\text{PDF}}}(\gamma, \sigma_C, \alpha_{ij}). 
\end{align}
The larger the additional error $\sigma_{\rm red}$ is, the less influence has the likelihood function \lcont and the more influence has the likelihood function \lpdf in the joint likelihood function. In the extreme case of $\sigma_{\rm red}\rightarrow\infty$, the likelihood \lcont would be ignored and we are using only the blue wavelength coverage of each spectrum with \lpdf as we have done in the main text. 
In the other extreme of $\sigma_{\rm red}=0$, we basically fix the continuum level to the best fit of the continuum model to the red side of each quasar spectrum due to the steepness of \lcont and are not able to regulate the continuum on the blue side via the PDF anymore. The likelihood function \lpdf would then only be used to estimate $\gamma$ and $\sigma_C$ with a fixed quasar continuum. 
A comparison of the two error terms $\sigma_{i, \rm noise}$ representing the Gaussian white noise that we add to each quasar spectrum with very high $\rm S/N$-ratio with
$\sigma_{\rm red}$ which reflects the limited ability to predict continuum on the blue side with an estimate of the continuum on the red wavelength side, reveals that the additional error $\sigma_{\rm red}$ dominates the error budget by roughly $2-3$ orders of magnitudes. 

\begin{figure}
\center
\includegraphics[width=.6\textwidth]{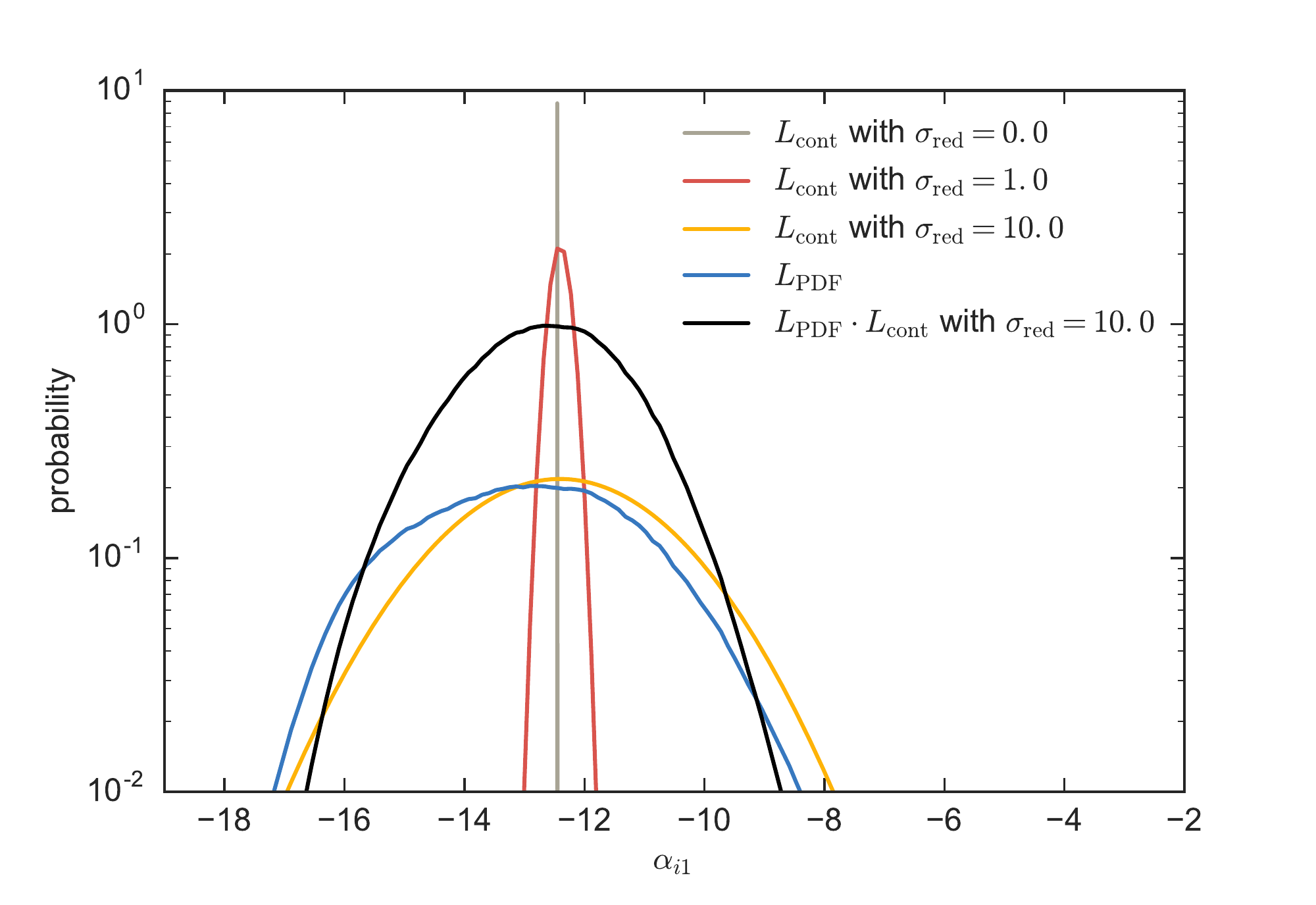}
\caption{Probability distributions given by the likelihood functions \lcont and \lpdf in dependency of the coefficient $\alpha_{i1}$ of the first principal component spectrum $\ket{\xi_1}$ for an example quasar $i$ at redshift $z=3$. The gray curve shows \lcont with no additional variance added to it, i.e. $\sigma_{\rm red}=0.0$. By increasing $\sigma_{\rm red}$ to $\sigma_{\rm red}=1.0$ the probability distribution becomes broader and less steep (red curve). For $\sigma_{\rm red}=10.0$ shown as the yellow curve the distribution is comparable in its steepness to the probability distribution of \lpdf (blue curve). The black curve shows the joint probability distribution function $\mathcal{L_{\text{PDF}}}\times\mathcal{L_{\text{cont}}}$ with $\sigma_{\rm red}=10.0$. \label{fig:different_L}} 
\end{figure}

We illustrate the differences in the steepness of the two likelihood
functions in Fig.~\ref{fig:different_L}. We show the normalized probability
distributions given by the the two different likelihood functions
\lcont and \lpdf dependent on one of the free parameters, the
coefficient $\alpha_{i1}$ of the first principal component spectrum
$\ket{\xi_1}$ for quasar $i$ at redshift $z=3$. All other principal
component coefficients $\alpha_{ij}$ with $j>1$ are thereby held
fixed. The blue curve shows the likelihood function \lpdf from the
\lya forest PDF, whereas the gray curve represents the likelihood
function \lcont from the continuum emission on the red side of the
quasar spectrum with $\sigma_{\rm red}=0.0$. 
The red and yellow curves show also the likelihood
function $\mathcal{L_{\text{cont}}}$, but now including an additional
error on the red wavelength side $\sigma_{\rm red}$. Setting
this additional error to $\sigma_{\rm red}=1.0$ (red curve) 
already causes the
likelihood \lcont to flatten and become considerably less steep. Its
probability distribution function widens compared to the gray curve with no additional error. Increasing $\sigma_{\rm
  red}$ will further flatten the likelihood function and thus increase
the width of the probability distribution. This is shown as the yellow
curve for $\sigma_{\rm red}=10.0$. This new likelihood function
results now in a probability distribution function that is of
comparable width to the one from the likelihood function \lpdf. Multiplying these two likelihood functions \lpdf and \lcont with $\sigma_{\rm red}=10.0$ results in the probability distribution shown in black. 

Note that the true value for this parameter for this particular quasar is $\alpha_{i1}\approx -11.4$, which is strongly disfavored by fitting the pixels redward of the \lya line if the $\sigma_{\rm red}$ is too small.

\subsection{Results Including the Red Wavelength Side of the Quasar Spectra}

\begin{figure*}
\centering
\includegraphics[width=.75\textwidth]{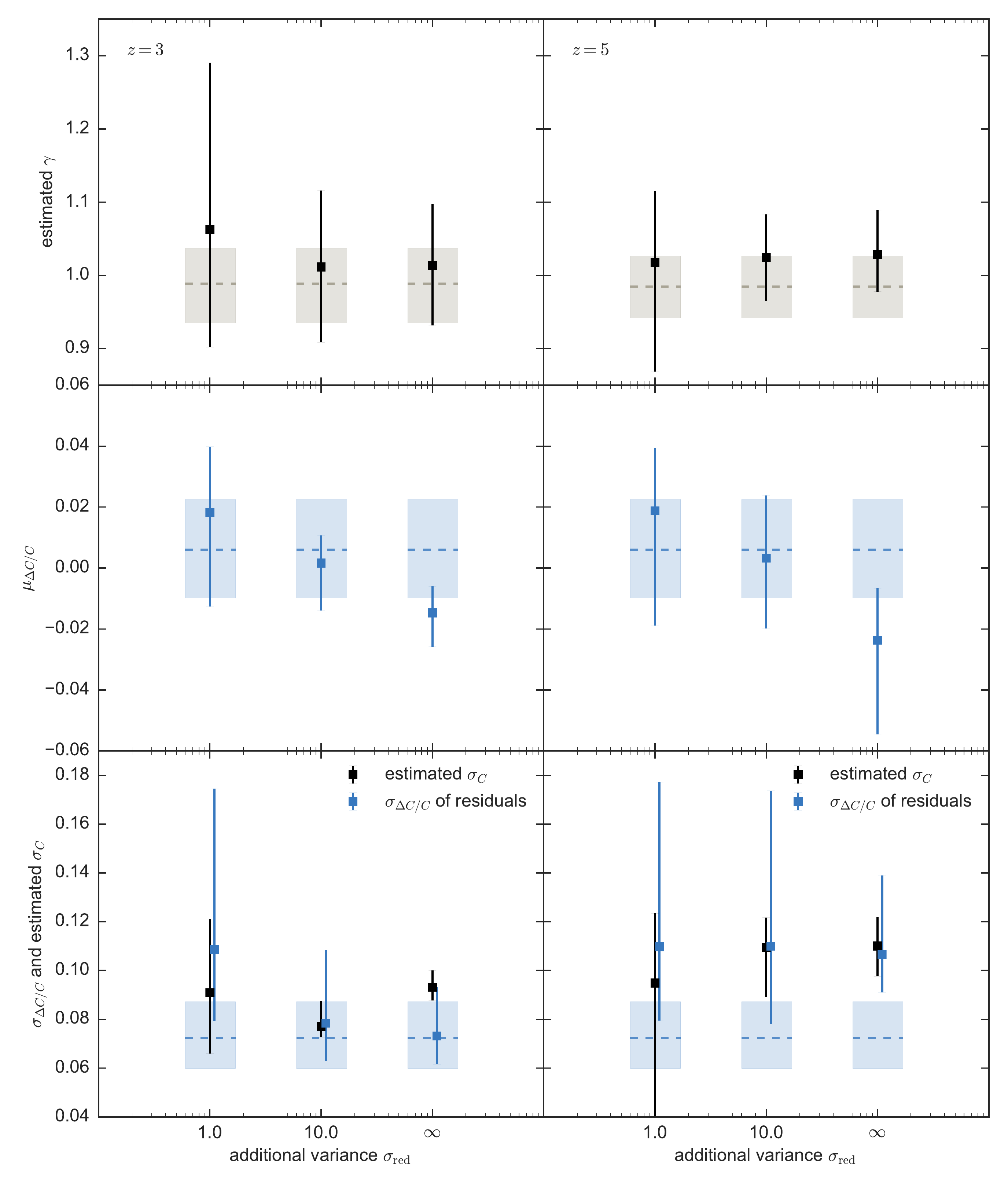}
\caption{Results of the analysis of the impact of the red wavelength side of each quasar continuum. The x-axis shows the additional error $\sigma_{\rm red}$ added to the spectra on wavelengths larger than \lya in order to make the two likelihood functions compatible. The upper panels show the median and $68\%$-region of estimates for $\gamma$. The gray areas indicate the precision achievable when estimating $\gamma$ for ensembles of noise-free quasar spectra and perfectly known quasar continua. The middle panels show the median and $68\%$-region of median values of the continuum flux residuals distributions in the \lya forest region. The blue shaded areas indicate the intrinsic precision of the chosen continuum model with $N_{\rm PCA}=2$. The lower panels show the uncertainty of the continuum estimation. The black data points show the median and $68\%$-region of estimates for the continuum error $\sigma_C$, whereas the blue data points show the actual measured values of the $16$th and $84$th percentiles of the continuum flux residuals, i.e. the actual PDF regulated continuum error $\sigma_{\Delta C/C}$. The blue shaded areas show again the intrinsic uncertainty of the chosen continuum model.}\label{fig:results_red}
\end{figure*}

In Fig.~\ref{fig:results_red} we show the results of our analysis when
including the information of each quasar spectrum redwards of the \lya
emission line. We show the same quantities as in the previous
Fig.~\ref{fig:results_inf} for $z=3$ and $z=5$ in the left and right
panels, respectively. The upper panels show the distribution of
medians of a hundred posterior probability distributions of $\gamma$
estimated from different realizations of an ensemble of ten quasar
spectra. The middle panels show the distribution of medians of the
continuum residual $\Delta C/C$ distributions of each run, and the
lower panels depict the continuum error --- once measured as the widths
of each continuum residual distribution, i.e. the actual PDF
regulated continuum error $\sigma_{\Delta C/C}$, shown as the blue
data points and once inferred via MCMC as a free parameter influencing
the shape of the PDF, i.e.  the estimated continuum error
$\sigma_C$. In this case the medians of the posterior probability
distributions are used again as the best estimate. The medians and
$68\%$-regions of this distributions of best estimates are shown as
the black data points.

However, the x-axis of Fig.~\ref{fig:results_red} now shows the
additional error $\sigma_{\rm red}$ that was added to the new
likelihood function $\mathcal{L_{\text{cont}}}$. A value of
$\sigma_{\rm red}\rightarrow\infty$ indicates that the likelihood
\lcont and thus the red side of each quasar spectrum has no influence
in the joint likelihood function and only the spectral coverage
bluewards of the \lya emission line is used to regulate the quasar
continuum. This is the case we have discussed extensively in \S~\ref{sec:res_all}.  Decreasing values of $\sigma_{\rm red}$ increase
the influence of \lcont in the joint likelihood function.  We run our
algorithm as before a hundred times with different realizations of an
ensemble of ten quasar spectra in order to avoid
statistical
fluctuations due to the limited number of quasar spectra. Throughout
this analysis we work with a continuum model with two PCS, i.e. $N_{\rm PCA}=2$. The gray and blue shaded
areas in all panels depict the same ideal cases as in
Fig.~\ref{fig:results_inf}, i.e. the gray shaded areas in the upper
panels show the precision on $\gamma$ achievable when assuming
noise-free spectra and perfectly known quasar continua and the blue
shaded areas in the middle and lower panels illustrate the intrinsic
uncertainty of the chosen continuum model.

The behavior of our algorithm at $z=3$ and $z=5$ is similar. Surprisingly,
the additional information of the wavelength coverage on the red side
of each quasar spectrum does not seem to improve the estimation of the
thermal parameter $\gamma$ significantly. The opposite even seems to
be the case at $z=3$ for values of $\sigma_{\rm red}$ that
are too small, i.e. $\sigma_{\rm red}=1.0$, where the $68\%$-region is significantly larger than for higher values of  $\sigma_{\rm red}$. 
However, the
additional information on the red wavelength side does improve the
quasar continuum estimation and reduces the small bias in
the medians of the continuum flux residual distributions (middle panels)
from $\mu_{\Delta C/C}\approx-1.5\%$ to $\mu_{\Delta C/C}\approx0.2\%$ at $z=3$ and from
$\mu_{\Delta C/C}\approx-2.3\%$ to $\mu_{\Delta C/C}\approx0.3\%$ at $z=5$ for 
$\sigma_{\rm red}=10.0$ compared to the previous case from \S~\ref{sec:res_all} when only
wavelengths bluer than \lya, i.e. $\sigma_{\rm
  red}\rightarrow\infty$, where taken into account.

The inferred estimates for the continuum error $\sigma_C$ (lower
panels) also reflect this improvement in the quasar continuum model for
$\sigma_{\rm red}=10.0$. At $z=3$ the measured (blue data
point) and inferred (black data point) estimates for the continuum
error agree perfectly and show that our algorithm results in a
reliable estimate of the actual continuum error. For lower values
of $\sigma_{\rm red}$ the continuum level estimation is set entirely by the very steep red side continuum likelihood, 
which causes the continuum
model on the blue wavelength side to not be a good fit anymore. Hence
the $68\%$-region of the estimates for the continuum error increases in the
actual PDF regulated continuum error $\sigma_{\Delta C/C}$
as well as in the inferred values for $\sigma_C$.  We mentioned
already that the agreement of the measured and estimated
continuum errors at $z=5$ is probably only coincidental
and reflects our choice of prior assumptions, since we do not have
good constraints on $\sigma_C$ at these redshifts from the \lya flux
PDF.

We conclude that the information of the spectral coverage on the red
wavelength side of each quasar spectrum does contain information
for improving the quasar continuum model on the blue wavelength
side of each spectrum. However, it is necessary to introduce an
additional error $\sigma_{\rm red}$ in order to prevent the
likelihood function \lcont to over-regulate the continuum model on the
red side of the quasar spectrum. We experimentally determined the best
value to be $\sigma_{\rm red}\approx 10.0$, i.e. we add an error
that enlarges the noise on the red wavelength side of each quasars by
multiple orders of magnitudes compared to the Gaussian noise that was
added to the spectra. Note that this choice for $\sigma_{\rm red}$ is purely based on the intent to match the widths of \lcont to $\mathcal{L_{\text{PDF}}}$, i.e. there is no physical meaning to its specific value. 

However, the additional information of the red spectral coverage, does \textit{not} improve the estimation of the thermal parameter $\gamma$, regardless of the exact choice for the value of $\sigma_{\rm red}$. Since the computational time increases significantly when adding the likelihood $\mathcal{L_{\text{cont}}}$, we advocate using it only when interested in the continuum estimation of the quasars. If the thermal parameters are the main parameters of interest, \lcont provides no additional information and the computational time will be significantly shorter without including the red side likelihood function.

\section{Justification of Our Analytical Model for the \lya Forest}

\subsection{Comparison to Hydrodynamical Simulations}

\begin{figure*}
\centering
\includegraphics[width=.75\textwidth]{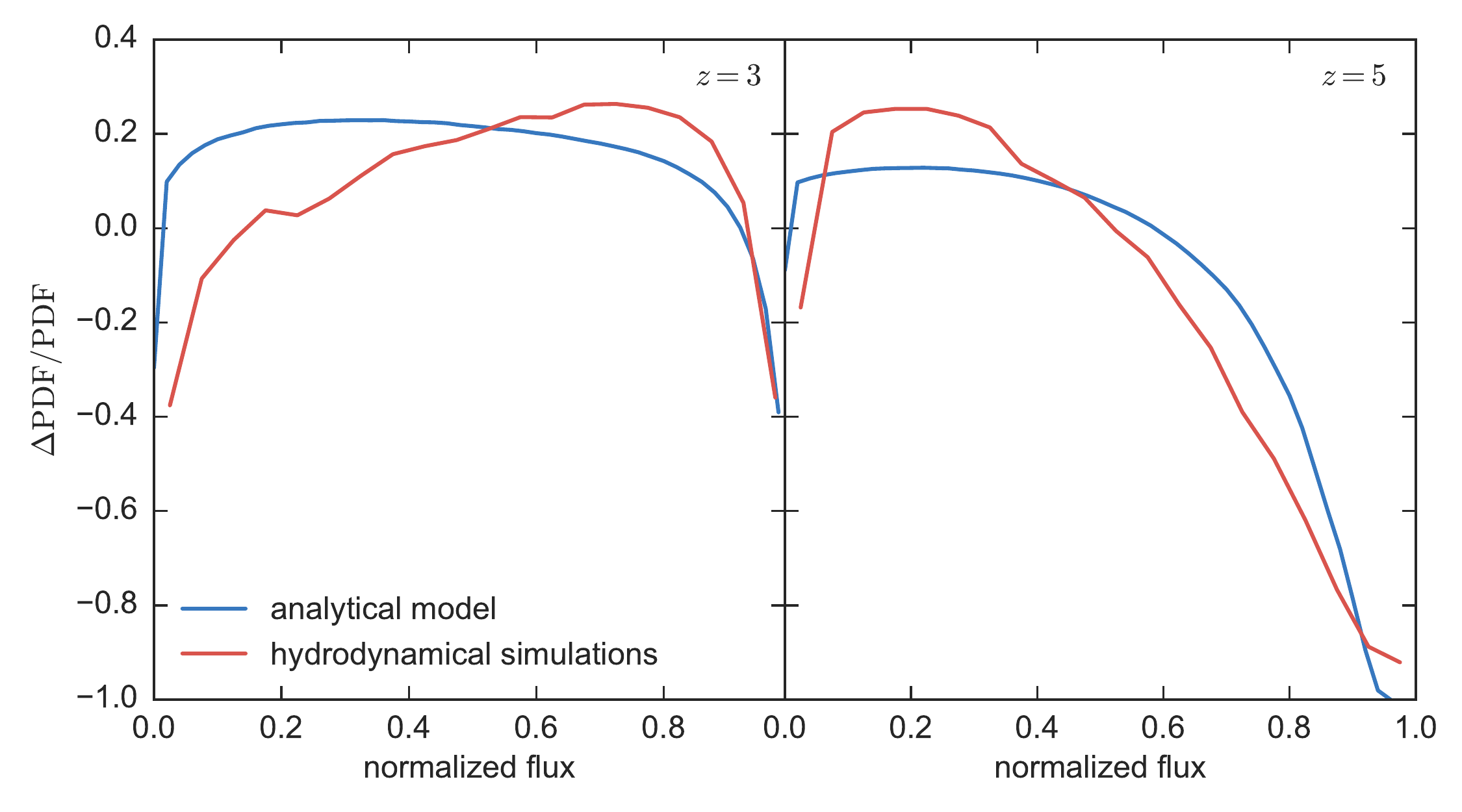}
\caption{Comparison between the flux PDF's from the analytical model for the \lya forest (blue curves) and from NyX hydrodynamical simulations (red curves) for $z=3$ and $z=5$ in the left and right panel, respectively. The y-axis is showing the relative change between a PDF models from eqn.~\ref{eq:pdf_change}}\label{fig:referee1}
\end{figure*}

This study considered a highly simplified analytical model to simulate the \lya forest absorption in high redshift quasars. We will show now that the flux PDF of our analytical model behaves qualitatively similar to the PDF form hydrodynamical simulations. Therefore we choose different runs from the NyX simulations \citep{Almgren2013}, that differ in the slope parameter $\gamma$ of the temperature-density relations, but other thermal parameters such as the temperature at mean density or the Jeans smoothing scale are kept fixed. The parameters of the chosen NyX simulations coming closest to our requirements had $\gamma=1.011$ ($T_0=10892$~K, $\lambda_P=64.5$~ckpc) and $\gamma=1.575$ ($T_0=10753$~K, $\lambda_P=66.5$~ckpc) at $z=3$, and $\gamma=0.9436$ ($T_0=7500$~K, $\lambda_P=69.66$~ckpc) and $\gamma=1.5219$ ($T_0=7713$~K, $\lambda_P=67.4$~ckpc) at $z=5$. Thus we compare these simulations to the PDF from the analytical models with $\gamma=1.01$ and $\gamma=1.58$ at $z=3$, and $\gamma=0.94$ and $\gamma=1.52$ at $z=5$. 
We calculate the difference in the PDF models for our analytical model and the hydrodynamical simulations and plot the relative change 
\begin{align}
\frac{\Delta\rm PDF}{\rm PDF} = \frac{\rm PDF (\gamma=1.58) - PDF (\gamma=1.01)}{\rm PDF (\gamma=1.01)},\label{eq:pdf_change} 
\end{align}
for $z=3$ in the left panel of Fig.~\ref{fig:referee1}, and correspondingly for $z=5$ in the right panel. The PDF from hydrodynamical simulations differs from the analytical model on a $10-20\%$ level, but shows qualitatively the same behaviour. Choosing a different smoothing kernel in our analytical model could reduce these differences (see \S~\ref{sec:kernel}), i.e. the analytical model with slightly higher smoothing kernel will resemble more closely the PDF from the NyX simulations. Thus we do not expect any discrepancies arising in the parameter estimation with our new method due to the the simplified analytical model with which the PDF models are constructed.

\subsection{Smoothing Kernel Applied to the Density Field}\label{sec:kernel}

\begin{figure*}
\centering
\includegraphics[width=.75\textwidth]{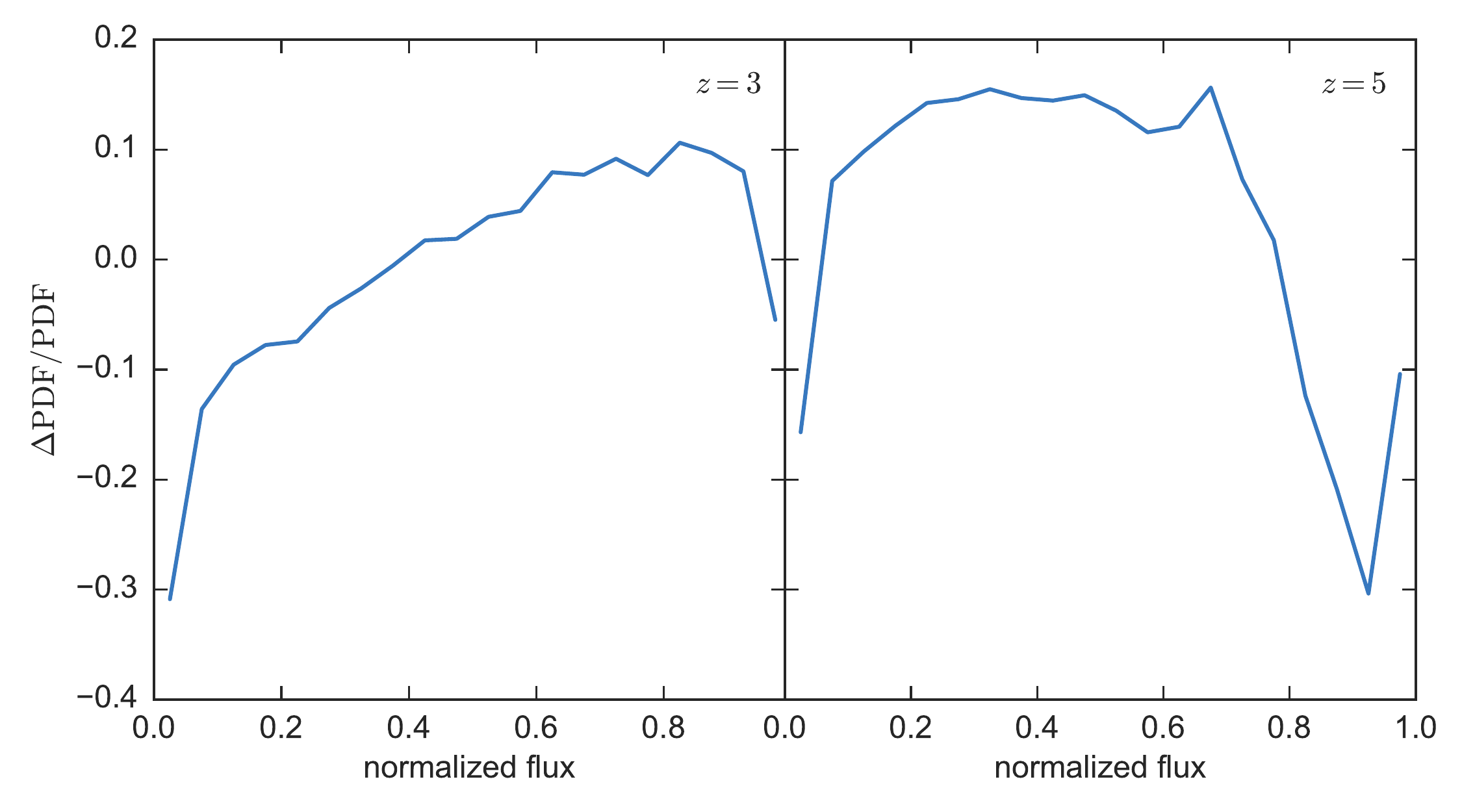}
\caption{Relative difference between two PDF models with smoothing kernels $\sigma_{\tau}=20$~km s$^{-1}$ and $\sigma_{\tau}\approx28$~km s$^{-1}$ for $z=3$ and $z=5$ in the left and right panel, respectively.}\label{fig:referee2}
\end{figure*}

In order to test the influence of the smoothing kernel $\sigma_{\tau}$ that has been applied to the density field (see \S~\ref{sec:mock_lya_forest}) we show in Fig.~\ref{fig:referee2} the relative differences between two PDF models with a smoothing kernel $\sigma_{\tau}=20$~km s$^{-1}$ and a smoothing kernel of $\sigma_{\tau}\approx28$~km s$^{-1}$, which roughly corresponds to a change in temperature of a factor of two, i.e. $\Delta T\approx 0.3$~dex. We plot
\begin{align}
\frac{\Delta\rm PDF}{\rm PDF} = \frac{\rm PDF (\sigma_{\tau}\approx 20~\rm km\,s^{-1}) - PDF (\sigma_{\tau}\approx 28~\rm km\,s^{-1})}{\rm PDF (\sigma_{\tau}\approx 20~\rm km\,s^{-1})}.  
\end{align}
Changing the smoothing kernel influences the shape of the PDF on a $10\%$ level, however, the slope parameter $\gamma$ has a significantly stronger influence on its shape. In the future, the smoothing kernel $\sigma_{\tau}$ could be easily incorporated into our PDF models and be treated as a nuisance parameter.

\end{document}